\documentstyle[12pt]{article}
\def\hybrid{\topmargin -20pt	\oddsidemargin 0pt
	\headheight 0pt	\headsep 0pt
	\textwidth 6.25in	
	\textheight 9.5in	
	\marginparwidth .875in
	\parskip 5pt plus 1pt	\jot = 1.5ex}

\hybrid

\begin{document}
\def\x{\times}
\def\beq{\begin{equation}}
\def\eeq{\end{equation}}
\def\beqa{\begin{eqnarray}}
\def\eeqa{\end{eqnarray}}
\def\D{ {\cal D}}
\def\L{ {\cal L}}
\def\C{ {\cal C}}
\def\N{ {\cal N}}
\def\calE{{\cal E}}
\def\lin{{\rm lin}}
\def\Tr{{\rm Tr}}
\def\mxth{\mathsurround=0pt }
\def\xversim#1#2{\lower2.pt\vbox{\baselineskip0pt \lineskip-.5pt
x  \ialign{$\mxth#1\hfil##\hfil$\crcr#2\crcr\sim\crcr}}}
\def\simgr{\mathrel{\mathpalette\xversim >}}
\def\simle{\mathrel{\mathpalette\xversim <}}

\def\R{ {\cal R}}
\def\I{ {\cal I}}
\def\calR{ {\cal R}}
\def\lag{Lagrangian}
\def\Kahler{K\"{a}hler}


\def\a{\alpha}
\def\b{\beta}
\def\dota{ {\dot{\alpha}} }
\def\lag{Lagrangian}
\def\Kahler{K\"{a}hler}
\def\kahler{K\"{a}hler}
\def\A{ {\cal A}}
\def\C{ {\cal C}}
\def\D{ {\cal D}}
\def\F{{\cal F}}
\def\L{ {\cal L}}
\def\R{ {\cal R}}
\def\x{ \times }
\def\beq{\begin{equation}}
\def\eeq{\end{equation}}
\def\beqa{\begin{eqnarray}}
\def\eeqa{\end{eqnarray}}
\def\S4{\frac{SO(4,2)}{SO(4) \otimes SO(2)}}
\def\P3{\frac{SO(3,2)}{SO(3) \otimes SO(2)}}
\def\MGd{\frac{SO(r,p)}{SO(r) \otimes SO(p)}}
\def\SOd{\frac{SO(r,2)}{SO(r) \otimes SO(2)}}
\def\SO2{\frac{SO(2,2)}{SO(2) \otimes SO(2)}}
\def\SUm{\frac{SU(n,m)}{SU(n) \otimes SU(m) \otimes U(1)}}
\def\SUS{\frac{SU(n,1)}{SU(n) \otimes U(1)}}
\def\SK{\frac{SU(2,1)}{SU(2) \otimes U(1)}}
\def\SU{\frac{ SU(1,1)}{U(1)}}

\renewcommand{\thesection}{\arabic{section}.}
\renewcommand{\theequation}{\thesection \arabic{equation}}



\sloppy
\newcommand{\be}{\begin{equation}}
\newcommand{\eq}{\end{equation}}
\newcommand{\ov}{\overline}
\newcommand{\un}{\underline}
\newcommand{\wt}{\widetilde}
\newcommand{\wh}{\widehat}
\newcommand{\p}{\partial}
\newcommand{\la}{\langle}
\newcommand{\ra}{\rangle}
\newcommand{\nl}{\newline}
\newcommand{\Nzahl}{{\bf N}  }
\newcommand{\zzahl}{ {\bf Z} }
\newcommand{\Zzahl}{ {\bf Z} }
\newcommand{\Qzahl}{ {\bf Q}  }
\newcommand{\Rzahl}{ {\bf R} }
\newcommand{\Czahl}{ {\bf C} }
\newcommand{\fs}[1]{\mbox{\scriptsize \bf #1}}
\newcommand{\ft}[1]{\mbox{\tiny \bf #1}}
\renewcommand{\arraystretch}{1.5}

\begin{titlepage}
\begin{center}

\hfill HUB-IEP-94/6\\
\hfill hep-th/9405002\\

\vskip .1in

{\bf MODULI SPACES AND
TARGET SPACE DUALITY SYMMETRIES IN $(0,2)\; Z_N$ ORBIFOLD THEORIES
WITH CONTINUOUS WILSON LINES }

\vskip .2in

{\bf Gabriel Lopes Cardoso,
Dieter L\"{u}st }  \\
\vskip .1in

{\em  Humboldt Universit\"at zu Berlin\\
Institut f\"ur Physik\\
D-10099 Berlin, Germany}\footnote{e-mail addresses:
GCARDOSO@QFT2.PHYSIK.HU-BERLIN.DE, LUEST@QFT1.PHYSIK.HU-BERLIN.DE}

\vskip .1in
and
\vskip .1in

{\bf Thomas Mohaupt}
\vskip .1in

{\em
DESY-IfH Zeuthen\\
Platanenallee 6\\
D-15738 Zeuthen, Germany}\footnote{e-mail address: MOHAUPT@HADES.IFH.DE}

\end{center}

\vskip .2in

\begin{center} {\bf ABSTRACT } \end{center}
\begin{quotation}\noindent
We present the
coset structure of the untwisted moduli space of heterotic
$(0,2) \;
Z_N$ orbifold compactifications with continuous Wilson lines.
For the cases where the internal 6-torus $T_6$ is given by the direct
sum $T_4 \oplus T_2$, we explicitly construct the
K\"{a}hler potentials associated with the underlying 2-torus $T_2$.
We then discuss the transformation properties of these K\"{a}hler
potentials under target space
modular symmetries. For the case where the $Z_N$ twist possesses
eigenvalues of $-1$, we find that holomorphic terms
occur in the
K\"{a}hler potential describing the
mixing of complex Wilson moduli.  As a consequence, the associated
$T$ and $U$ moduli are also shown to mix under target space modular
transformations.

\end{quotation}
April 1994\\
\end{titlepage}
\vfill
\eject

\newpage

\section{Introduction}

\hspace*{.3in} Orbifold
models \cite{DHVW1,DHVW2,NSV1,NSV2,IbaNilQue2,IbaNilQue}, describing the
compactification of the heterotic
string from ten dimensions down to four, have
been extensively studied in the past due to the fact
that these models are exactly solvable and that they can predict
semi-realistic
physics.
Orbifold compactifications
possess various continuous parameters, called moduli,
corresponding
to marginal deformations of the underlying
conformal theory. They enter into the 4D $\N = 1$ supersymmetric
low-energy effective Lagrangian
as chiral matter fields with flat potentials.  Moduli take their
values in a manifold called moduli space.  For models yielding
$\N =1$ space-time supersymmetry
this moduli
space is, locally, a K\"{a}hlerian manifold.  Its K\"{a}hler potential
is one of the three functions \cite{CreJul,CreFer}
which describe the coupling of moduli
fields to $\N=1$
supergravity in the 4D low-energy effective Lagrangian.
It has been known for quite some time now that
a certain class of moduli, called continuous
Wilson lines, can occur whenever the gauge twist in the $E_8 \otimes
E_8$ root lattice is realized not by a shift, but rather by a rotation
\cite{IbaNilQue,IbaMasNilQue,Moh2}.  Continuous Wilson lines are of big
interest, not only because they enter into
the 4D low-energy effective Lagrangian as described above, but also
because turning them on leads to $(0,2)$ models with
observable gauge groups
smaller than the generic gauge group $E_6 \otimes H$ (where
$H=SU(3),SU(2) \otimes
U(1),U(1)^2$)
occuring in
$(2,2)$ symmetric $Z_N$ orbifold compactifications.
This gauge symmetry breaking
is related to the stringy Higgs effect \cite{IbaLerLu}.

It was
only recently \cite{Moh2},
however, that a first step towards a complete classification
of the untwisted moduli space of $Z_N$ orbifold theories
with continuous Wilson lines was taken.
After reviewing some of the relevant facts about toroidal and orbifold
compactifications in section 2, we will in section 3
derive the local structure of the untwisted
moduli space of asymmetric
$Z_N$ orbifolds with continuous Wilson lines.  We find
that its local structure is given by a direct product of $\SUm$ and $\MGd$
cosets and that it is
entirely determined by the eigenvalues
of the twist $\Theta$ on the underlying Narain lattice.
We then proceed with a general discussion of
target space modular symmetries in asymmetric $Z_N$ orbifolds with
continuous Wilson lines.  This is presented in section 4.
Target space duality symmetries consist
of discrete reparametrisations of the moduli fields which change the
geometry of the internal space but leave the underlying conformal field
theory invariant.  This implies that certain
points in the moduli space of orbifold models have to be
identified.  Thus, the moduli space of the underlying conformal
field theory is an orbifold and not a smooth
manifold. In section 4,
we also introduce two sets of standard coordinates on the
$\MGd$ cosets, namely real homogenous and real projective coordinates.
This is useful because the
group of modular transformations acts in a simple way on these coordinates.
Next, we especialise to the case of $(0,2)$
$Z_N$ orbifold compactifications with continuous Wilson lines yielding
$\N=1$ space-time supersymmetry.  We then proceed, in section 5,
to explicitly
construct the K\"{a}hler potentials for some of these moduli spaces.
Namely, we will focus on
the $Z_N$ orbifold compactifications
for which the internal 6-torus $T_6$ can be decomposed
into a direct sum $T_4 \oplus T_2$.
Then, by using well known techniques \cite{PorZwi,FerPor,FerKouLuZw},
we will derive
the K\"{a}hler potentials for the moduli spaces associated with the
underlying 2-torus $T_2$.
For the case
when the twist operating on the internal $T_2$ torus has eigenvalues $-1$
we find that, in the presence of $r-2$ complex Wilson lines,
the associated coset $\SOd$ doesn't factorise anymore into two submanifolds,
contrary to the
$\SO2$ case with
no Wilson lines turned on.  Moreover, we find that
the associated K\"{a}hler potential contains a holomorphic term describing
the mixing of complex Wilson lines.  Such a term is precisely of the type
which recently has been shown \cite{LouKap,Anto}
to induce a mass term for Higgs particles of the same order of the
gravitino mass
once supergravity is spontaneoulsy broken at low energies.  In section
6 we proceed to explicitly
discuss target space modular symmetries \cite{FLST}
of the K\"{a}hler potentials
constructed in section 5.
We show that, for the K\"{a}hler potentials we have explicitly
constructed,
these
discrete reparametrisations induce particular K\"{a}hler transformations
on the K\"{a}hler potentials.  Hence, these target space duality
transformations are
symmetries of the 4D $\N=1$ tree level
low-energy Lagrangian \cite{FLST}.
These target-space duality symmetries also
manifest themselves in the string threshold corrections that are of importance
for the unification of gauge couplings
\cite{Kaplu,Dix,AntNar,AntGav,DFKZ,AELN,ILR,IbaLu,May,BLST1,BLST2}.
  We point out
that, for the case where the twist operating on the internal $T_2$
has eigenvalues of $-1$, the associated $T$ and $U$ moduli mix under
target space duality transformations due to the presence of the mixing
terms between complex Wilson lines in the K\"{a}hler potential. We
present
our conclusions in section 7.

\setcounter{equation}{0}
\section{Toroidal and Orbifold Compactifications}

\hspace*{.3in}Let us
first briefly recall some of the relevant facts about toroidal
compactifications with general constant background fields
\cite{Nar,NSW}.
If one compactifies the ten--dimensional heterotic $E_{8} \otimes E_{8}$
string on a $d$--dimensional torus,
\be     {\bf T}^{d} = \frac{ {\bf R}^{d} }{\Lambda}
\eq
(where $\Lambda$ is a $d$--dimensional lattice)
then the moduli dependent degrees
of freedom can be parametrized by $16 + 2 d$ chargelike integer
quantum numbers, namely the winding numbers $n^{i}$, the
internal momentum numbers $m_{i}$ and the charges $q^{A}$ of the
leftmoving current algebra which is generated by the extra leftmoving
sixteen coordinates ($i=1,\ldots,d$, $A=1,\ldots, 16$).

The moduli dependence of the untwisted states is encoded in the
$16 + 2d$ dimensional Narain lattice $\Gamma$. If one expands
the Narain vector ${\bf P}$ of an untwisted state in terms of a standard
lattice basis \cite{Gin} $k^{i}, \ov{k}_{i}, l_{A}$, then
the quantum numbers appear as components, whereas the moduli
dependence is absorbed into the geometry of the lattice:
\be    {\bf P} = q^{A} l_{A} + n^{i}  \ov{k}_{i} +
                 m_{i} k^{i} \in \Gamma
\eq
The moduli are usually grouped into a symmetric matrix $G_{ij}$,
which denotes the lattice metric of the $d$--dimensional lattice
$\Lambda$, an antisymmetric matrix $B_{ij}$ and $d$
sixteen dimensional vectors ${\bf A}_{i}$, called Wilson lines.
The standard basis of $\Gamma$ can then be constructed in
terms of a basis ${\bf e}_{A}$ of the $E_{8} \otimes E_{8}$
lattice and of bases ${\bf e}_{i}$, ${\bf e}^{i}$ of $\Lambda$
and $\Lambda^{*}$ (the dual of $\Lambda$) as a function of
the moduli $G_{ij}, B_{ij}, {\bf A}_{i}$ \cite{Gin}:
\be    k^{i} = \left(
       {\bf 0}, \frac{1}{2} {\bf e}^{i};\frac{1}{2} {\bf e}^{i} \right)
\eq
\be    \ov{k}_{i} = \left(
       {\bf A}_{i}, (G_{ij} + B_{ij} - \frac{1}{4} ({\bf A}_{i} \cdot
         {\bf A}_{j}) ) {\bf e}^{j};
                    (-G_{ij} + B_{ij} - \frac{1}{4} ({\bf A}_{i} \cdot
        {\bf A}_{j}) ) {\bf e}^{j} \right)
\eq
\be     l_{A} = \left(
        {\bf e}_{A}, -\frac{1}{2} ({\bf e}_{A} \cdot {\bf A}_{i})
        {\bf e}^{i};
        -\frac{1}{2} ({\bf e}_{A} \cdot {\bf A}_{i}) {\bf e}^{i}
        \right)
\label{lA}
\eq
Defining the $16 \times d$ matrix $A_{Ai}$ as
\be     A_{Ai} = {\bf e}_{A} \cdot {\bf A}_{i}
\label{DefA}
\eq
yields
\beq
{\bf A}_i \cdot {\bf A}_j = C^{AB} A_{Ai} A_{Bj}
\label{AiAj}
\eeq
where the metric $C_{AB}= {\bf e}_A \cdot {\bf e}_B$ for lowering
and raising $A$-indices denotes the Cartan metric of the
$E_8 \otimes E_8$ lattice.
Another parametrization of the moduli, which will turn out to be
quite useful later on, is given by a $d \times d$ matrix $D_{ij}$
defined by
\be     D_{ij} = 2(B_{ij} - G_{ij} - \frac{1}{4}({\bf A}_{i} \cdot
        {\bf A}_{j}))
\label{DefD}
\eq
At those points in the moduli space, where one of the matrices
(\ref{DefA}) and (\ref{DefD})
becomes integer valued, both the symmetry of the Narain lattice
and the gauge symmetry of the model are enhanced \cite{Moh1}.

Another useful representation of the Narain vector is
to specify its components with respect to an orthonormal frame,
which allows one to separate the $16 + d$ leftmoving from the
$d$ rightmoving degrees of freedom
\be    {\bf P} = ( {\bf P}_{L}; {\bf P}_{R} )
\eq
In terms of this decomposition the condition
$(L_{0} - \tilde{L}_{0}) | \Phi \ra = 0$ for physical states
reads
\be \frac{1}{2} ( {\bf P}_{L}^{2} -  {\bf P}_{R}^{2} )
     + N + \wt{N} - 1 = 0
\eq
Since the (moduli independent) contribution of the number operators
$N$ and $\wt{N}$ is an integer\footnote{This is true
after applying the GSO condition and after absorbing the normal ordering
constant of the NS sector into the definition of the rightmoving
number operator.} it is evident that the Narain lattice must be
an even lattice with respect to the indefinite bilinear form of
type $(+)^{16 + d} (-)^{d}$. As shown by Narain \cite{Nar}
modular invariance
implies that the lattice $\Gamma$ must also be selfdual. Since even selfdual
lorentzian lattices are unique up to isometries, this then implies
that the allowed deformations of such a lattice $\Gamma$ form a group
isomorphic to $O(16 + d,\; d)$.

The moduli dependent contribution to the mass $M$ of an untwisted state is
given by
\be \alpha' M^{2} = ({\bf P}_{L}^{2} + {\bf P}_{R}^{2})
+ \cdots
\label{m15}
\eq
Since not only the spectrum but also the interactions are
invariant under the subgroup $O(16+d) $$\otimes O(d)$
of $O(16+d, \; d)$, the moduli space of toroidal compactifications
is locally given by the coset space \cite{Nar}
\be  {\cal M}_{T} \simeq \frac{ O(16+d, \; d) }{ O(16 + d) \otimes
  O(d) }
\label{ModuliSpaceT}
\eq
In order to
get the global geometry one has to take into account further
discrete identifications due to duality (also called modular) symmetries
of the target space, which will be discussed later.

Toroidal
compactifications
are, however,
not of big phenomenological interest, because they all yield models
with
an extended ${\cal N} = 4$ space--time supersymmetry, which doesn't
admit chiral matter multiplets, and with gauge groups of
rank $16 + d$ \cite{Nar}.
They are, nevertheless,
the natural starting point for the construction
of more realistic models, namely orbifold models.


It is well known from the work of Dixon, Harvey, Vafa and Witten
\cite {DHVW1,DHVW2}
that by modding out rotations both the number of space--time
supersymmetries and the rank of the gauge group can be reduced.
If one starts with a toroidal compactification
these rotations
must be automorphisms of finite order of the corresponding
Narain lattice $\Gamma$ \cite{NSV1}.
We will study the case in which
the point twist group ${\cal P}$ defining the orbifold
is a cyclic group
\be {\cal P} = \la \Theta \ra = \{ \Theta, \Theta^{2}, \ldots,
                \Theta^{N} = {\bf 1} \}
\eq
generated by a single twist $\Theta$ satisfying
\be   \Theta \in \mbox{AUT}(\Gamma),\;\;\;
      \Theta^{N} = {\bf 1}
\eq
As shown by Narain, Sarmadi and Vafa \cite{NSV2}
holomorphic factorization and modular invariance imply that
the twist must not mix left- and rightmoving degrees of freedom.  It
must
therefore be a rotation\footnote{
The asymmetric orbifold construction given in \cite{NSV1}
is slightly more general since it also allows for the
modding out of a
rotation followed by a translation.
Note that modding out by translations is
much simpler and better understood as it is equivalent to imposing
different toroidal boundary conditions. For the purpose of
trying  to learn more
about the effect of modding out rotations, we will keep the situation
as simple as possible and, in the following, only consider pure
rotations.} (not just a pseudo--rotation)
\be  \Theta = \Theta_{L} \otimes \Theta_{R} \in
     O(16 + d) \otimes O(d)
\label{m1}
\eq
We will, in the next section, determine the local structure
of the moduli spaces for orbifolds defined by a twist as given in
(\ref{m1}).
Since most of the work on orbifolds has, up to now, focused on more special
constructions we will, however, first have to
recall some more facts and results.

People have, from the
beginning, been especially interested in orbifold
models that can be interpreted as compactifications on a
six--dimensional orbifold \cite{DHVW1,DHVW2}.
In these cases the twist $\Theta$ of the Narain lattice $\Gamma$
must act
in a left-right symmetric way, to be specified below, so as
to have well defined
coordinates
on the internal $d$--dimensional orbifold.
More precisely, the twist $\Theta$ must be given in terms of
a $d$--dimensional twist $\theta$ which defines this orbifold and
an additional gauge twist $\theta'$ which is an automorphism of
the $E_{8} \otimes E_{8}$ root lattice.  That is,
if one decomposes the Narain vector as
\be {\bf P} = (p^{A}, p^{i}_{L}; p_{R}^{i} )
\eq
then the twist $\Theta$ must be given as
\be \Theta = \theta' \otimes \theta \otimes \theta
           \in O(16) \otimes (O(d) \otimes O(d) )_{\mbox{diag}}
\label{m2}
\eq
We will refer to all compactifications, for which the
twist $\Theta$ is given by (\ref{m2}), as
{\em orbifold compactifications}.  Note that (\ref{m2}) is a special
case of (\ref{m1}).

One further restriction that is often used is to consider only
Narain lattices of the special form $\Gamma_{16} \oplus \Gamma_{6;6}$
where $\Gamma_{16}$ denotes the root lattice of $E_{8} \otimes E_{8}$.
This means that most of the deformation parameters, namely
the $16 \cdot d$ parameters
corresponding to the Wilson lines $A_{Ai}$,
are set to zero.
One can then replace the gauge twist $\theta'$ by an equivalent
shift (i.e. by a translation) which is much easier to handle.
However, the price of this simplification is quite high, as
the rank of the gauge group is then at least\footnote{
The rank of the gauge group cannot be reduced by shifts
but only by rotations. More precisely, the rank of the
gauge group of an asymmetric orbifold is greater or equal to
the number of nontrivial
eigenvalues of $\Theta_{L}$, because for each eigenvalue 1 there
is a twist invariant leftmoving
oscillator and, therefore, an unbroken $U(1)$.} 16. Although it is
possible to have nonvanishing
Wilson lines when using the shift realization, they are then
constrained to a discrete set of values and, hence,
are not moduli of the orbifold model \cite{IbaNilQue2,IbaMasNilQue}.
Since discrete Wilson lines
act like additional shifts, they also cannot reduce
the rank of the gauge group but only break (or extend) the gauge group.
On the other hand, it was pointed out in
\cite{IbaNilQue,IbaMasNilQue}
that, if one realizes the gauge twist by a rotation,
some of the components of the Wilson lines are still moduli and that
they can be used to reduce the rank of the gauge group
below 16.
Thus, it is important to keep
the continuous Wilson lines in the game and we will do so in the
following.

Clearly, a deformation of the Narain lattice can only lead to
a modulus of an orbifold model if the twist $\Theta$ is still
an automorphism of the deformed lattice. This was used in
\cite{IbaMasNilQue} to derive a set of
equations for the moduli which, in
principle, allow one to decide which of the toroidal moduli
are still moduli of the orbifold model and which are frozen to
discrete values. In \cite{ErlJunLau,Jun} it was shown how these
equations can be explicitly solved for the moduli in the case
of bosonic
or heterotic orbifold compactifications without Wilson lines.
For the number of surviving $G_{ij}$ and $B_{ij}$ moduli closed
formulas were derived. This was later \cite{Moh2}
generalized to heterotic
orbifold compactifications with continuous Wilson lines.
One drawback of the approach used in \cite{Moh2}
is that one can derive the number of
moduli, but the expected coset structure of the moduli space remains
obscure. In the case of vanishing Wilson lines this coset structure
was derived in \cite{FKP,CveLouOvr}
for all the ${\bf Z}_{N}$ orbifold compactifications
with ${\cal N} = 1$ and  ${\cal N} = 2$ space--time
supersymmetry. In that approach one uses
symmetries of the world sheet action to constrain the
K\"ahler potential appearing
in the 4D effective action. The associated coset is then
obtained from the explicit expression of the
K\"ahler potential. In the next section
we will use a different method for determining the local structure
of the moduli space of asymmetric $Z_N$ orbifolds.
We will not make use of the effective
action, but rather of
the compatiblity equation between the Narain twist
$\Theta$ and the moduli.  Note that we will be dealing with orbifolds
defined by a twist $\Theta$ as given in (\ref{m1}).

\setcounter{equation}{0}

\section{The coset structure of asymmetric ${\bf Z}_{N}$ orbifolds}

\hspace*{.3in}
Consider the coset (\ref{ModuliSpaceT})
which parametrizes the moduli space of toroidal compactifications
locally.
The simplest way of arriving at the untwisted
moduli space of a general ${\bf Z}_{N}$
orbifold (locally) is simply to find the subspace of this coset that
is compatible with the action of the twist $\Theta$ on the underlying
Narain lattice $\Gamma=$ $\Gamma_{16 + d;d}$.

Suppose now that $\Gamma$ is a lattice on which $\Theta$ acts as an
automorphism. A deformation ${\cal T}
 \in O(16+d,\;d)$ of $\Gamma$ is compatible
with $\Theta$ if and only if $\Theta$ is also an automorphism of the
deformed lattice $\Gamma' = {\cal T} (\Gamma)$. But this is, by inspection,
equivalent
to ${\cal T}^{-1} \Theta {\cal T}$ being in the point group $\cal{P}$
of $\Gamma$.
Since we are taking\footnote{There may, of course, exist lattices
which are more symmetric than required when
modding out by $\Theta$
and, therefore, have bigger point groups. It is possible
to define orbifolds by modding out these bigger groups. The number
of allowed deformations will then be different.
For our purpose these models are just a subset of models
with extended symmetry because we want to find all lattices
whose point {\em symmetry} group contains the cyclic group
generated by $\Theta$, which then is chosen to be the
point {\em twist} group.}
the point group $\cal P$ to be the cyclic group generated
by $\Theta$, this then means that
\be    {\cal T}^{-1} \Theta {\cal T} = \Theta^{k} \;\;,
1 \leq k <N
\label{m13}
\eq
for some $k$.
That is, $\cal T$ is in the {\em normalizer}
$\cal{N}$
of the point group $\cal P$ in $O(16+d,\;d)$
\be  {\cal T} \in {\cal N} ( {\cal P}, O(16+d,\; d) )
\label{m14}
\eq
Statement (\ref{m14}), namely that $\cal T$ has to be in
the normalizer $\cal N$ of the point group, also
holds for bigger (abelian or non-abelian)
point groups $\cal P$ which have more than one generator.
It is, though,
intuitively clear that a twist $\Theta$ and a deformation
$\cal T$
do not
have to strictly
commute, but that they have to commute
on orbits, that is up to
point transformations as in (\ref{m13}).

Of course, only those deformations $\cal T$
with $k=1$ can be continuously
connected to the identity, whereas the others will describe
nontrivial, discrete deformations. This corresponds to
the appearence of discrete background fields in the
standard approach \cite{Erl}. On the other hand, any special
solution of equation (\ref{m13})
with $k \not= 1$ can be continuously
deformed by any solution to (\ref{m13}) with
$k = 1$. This means that, in order to identify the moduli,
one has to
find the general solution to (\ref{m13}) with $k=1$.

We will, therefore, in the following
only deal with the (most general) case of purely
continuous background fields and set $k=1$.
After introducing matrices with respect to an orthonormal
basis of ${\bf R}^{16+d, \; d}$ we have to solve
the homogenous matrix equation
\be [\Theta, {\cal T}] = \mbox{\bf 0}
\label{comequ}
\eq
for\footnote{
We use the same characters for the maps $\Theta$ and $\cal T$
themselves and for the matrices representing them.} $\cal T$.
We proceed to show
that the moduli space of this equation only
depends on the eigenvalues of $\Theta$. The method used in the
following is a modification of the method used in \cite{FerFreSor}
for ${\bf Z}_{3}$ orbifold compactifications without Wilson lines.
First recall that
$\Theta$ must be a proper rotation (\ref{m1}),
that is, it must be an element of
$O(16+d) \otimes O(d)$.
The eigenvalues of the twist are $N$--th roots of unity. Those which
are real must be equal to $\pm 1$, whereas the complex ones come in
pairs of complex conjugated numbers of length 1. Let us denote
the number of eigenvalues $\pm1$ in the left (right) part of $\Theta$
by $d_{\pm 1}^{(L)}$ ($d_{\pm 1}^{(R)}$) and the total number by
$d_{\pm 1}$ $=$ $d_{\pm 1}^{(L)} + d_{\pm 1}^{(R)}$. Analogously
the number of complex pairs $\exp(\pm i \phi_{k})$ of eigenvalues
of the left (right) part is denoted by $p_{k}$ ($q_{k}$).

By relabeling the orthonormal basis of $\mbox{\bf R}^{16+d,\;d}$,
the matrix of the twist $\Theta$ can be brought to the form
\be
\left( \begin{array}{cccccc}
\begin{array}{cc}
R_{1} & 0 \\ 0 & R_{1}' \\
\end{array}
& {\bf 0} & \cdots & \cdots & \cdots & {\bf 0} \\
{\bf 0} & \ddots & & & & \vdots \\
\vdots & &
\begin{array}{cc}
R_{i} & 0 \\ 0 & R_{i}' \\
\end{array}
& & & \vdots \\
\vdots & & & \ddots & & \vdots \\
\vdots & & & &
\begin{array}{cc}
- {\bf 1}_{d_{-1}^{(L)} } & 0 \\ 0 & - {\bf 1}_{d_{-1}^{(R)}} \\
\end{array}
& {\bf 0} \\
{\bf 0} & \cdots & \cdots & \cdots & {\bf 0} &
\begin{array}{cc}
{\bf 1}_{d_{1}^{(L)} } & 0 \\ 0 &  {\bf 1}_{d_{1}^{(R)}} \\
\end{array}
\\
\end{array} \right)
\eq
where
\be
R_{i} = \left( \begin{array}{cc}
c_{i} {\bf 1}_{p_{i}} & -s_{i} {\bf 1}_{p_{i}} \\
s_{i} {\bf 1}_{p_{i}}  & c_{i} {\bf 1}_{p_{i}}  \\
\end{array} \right) \;\;,\;\;
R_{i}' = \left( \begin{array}{cc}
c_{i} {\bf 1}_{q_{i}} & -s_{i} {\bf 1}_{q_{i}} \\
s_{i} {\bf 1}_{q_{i}}  & c_{i} {\bf 1}_{q_{i}},  \\
\end{array} \right)
\eq
with
\be
c_{i} = \cos(\phi_{i}) \;\;,\;\; s_{i} = \sin(\phi_{i})
\eq
 Then the matrix of an admissible deformation $\cal T$
with respect to the
same basis has the blockdiagonal form
\be
\left( \begin{array}{cccccc}
T_{1} & {\bf 0} & \cdots & \cdots & \cdots & {\bf 0} \\
{\bf 0} & \ddots &  & & & \vdots \\
\vdots & & T_{i} & & & \vdots \\
\vdots & & & \ddots & & \vdots \\
\vdots & & & & P & {\bf 0} \\
{\bf 0} & \cdots & \cdots & \cdots & {\bf 0} & Q \\
\end{array} \right).
\eq
Since $\cal T$ $\in$ $O(16 + d,\;  d)$
we get that
\be
T_{i} \in O(2p_{i},\;  2q_{i} ), \;\;\;
P \in O(d_{-1}^{(L)},\; d_{-1}^{(R)}), \;\;\;
Q \in O(d_{1}^{(L)},\; d_{1}^{(R)} )
\eq
Whereas the matrices $P$ and $Q$ are not further constrained by
the commutator equation (\ref{comequ}), the $T_{i}$ must commute
with the twist matrix restricted to the $i$--th complex eigenspace.
Decomposing $T_{i}$ into suitable blocks as
\be
      T_{i} =
      \left( \begin{array}{cc}
      A & B \\ C & D
      \end{array} \right)
\eq
yields (3.3) as
\be   \left[ \left( \begin{array}{cc}
      R_{i} & 0 \\ 0 & R_{i}'
      \end{array} \right),
      \left( \begin{array}{cc}
      A & B \\ C & D
      \end{array} \right) \right] = \mbox{\bf 0}
\label{comeq}
\eq
The blocks $A,B,C,D$ depend of course on the index $i$, but since
the different eigenspaces decouple it is possible and convenient to
suppress this label.

Equation (\ref{comeq}) implies that
$R_{i} A = A R_{i}$ for the
$2p_{i} \times 2p_{i}$ block $A$.  Similar
equations hold for $B, C$ and $D$.
These equations can now again be analyzed blockwise. In the case of
$A$ (\ref{comeq}) gives
\be  [A, R_{i}] =
     \left[
     \left( \begin{array}{cc}
     A_{1} & A_{2} \\ A_{3} & A_{4} \\
     \end{array} \right),
     \left( \begin{array}{cc}
     c_{i} \mbox{\bf 1}_{p_{i}} & -s_{i} \mbox{\bf 1}_{p_{i}} \\
     s_{i} \mbox{\bf 1}_{p_{i}} & c_{i} \mbox{\bf 1}_{p_{i}} \\
     \end{array} \right) \right] = \mbox{\bf 0}
\eq
implying that only two of the four $p_{i} \times p_{i}$ blocks $A_{i}$
are independent, namely
\be   A = \left( \begin{array}{cc}
      A'& -A''  \\ A''& A' \\
      \end{array} \right)
\eq
where
$A' = A_{1} = A_{4}, \;\; A''= -A_{2} = A_{3}$.
The other blocks $B, C, D$
of $T_{i}$ have the same structure. The off--diagonal
blocks $B$ and $C$ are, however, in general not quadratic.

The matrices $T_{i}$ form a group called the centralizer of
$\Theta$ (restricted to the $i$--th eigenspace) in $O(2p_{i}, \; 2q_{i})$.
The special structure found for these matrices resembles the one
appearing in the
standard isomorphism between $GL(n, \mbox{\bf C})$ and a
$2n^{2}$ dimensional subgroup of $GL(2n, \mbox{\bf R})$ given by
\be    GL(n, \mbox{\bf C}) \ni
       m = m' + i m'' \longleftrightarrow
       M = \left( \begin{array}{cc}
           m' & -m'' \\ m'' & m' \\
        \end{array} \right)
       \in GL(2n, \mbox{\bf R})
\label{standaut}
\eq
The only modification needed is a permutation of certain rows and
columns in $T_{i}$, in order to reposition some of the blocks. Since
such a permutation is
an automorphism of $GL(2p_{i} + 2 q_{i})$ the composition
with (\ref{standaut}) yields again an isomorphism

\be T_{i}          = \left( \begin{array}{cccc}
                    A' & -A'' & B' & - B''\\
                    A'' & A' &B''  & B'\\
                    C' & - C'' &  D' & - D'' \\
                    C'' &  C' & D'' & D' \\
                   \end{array} \right)
\longleftrightarrow     \left( \begin{array}{cccc}
                    A'& B' &-A'' & -B'' \\
                    C'& D' &-C'' & -D'' \\
                    A''& B'' & A'& B' \\
                    C''&D''  & C'& D' \\
                    \end{array} \right)
\longleftrightarrow
\eq
\be
\longleftrightarrow     \left( \begin{array}{cc}
                    A' + i A'' & B' + i B''\\
                   C' + i C'' & D' + i D'' \\
                   \end{array} \right)
                    = \left( \begin{array}{cc}
                    a & b \\ c & d \\
                    \end{array} \right) = t_{i}
\eq
Since the $T_{i}$ must not only commute with the twist but also
be in $O(2p_{i},\; 2q_{i})$, we finally have to translate this
into a condition for the $t_{i}$.
The pseudo--orthogonality\footnote{The condition on the determinant
does not lead to a relation between the matrix elements of
$T_{i}$, because the determinant of a pseudo--orthogonal
matrix can only take discrete values.}
 of $T_{i}$ can be expressed in terms
of the blocks $A,B,C,D$ as
\be \begin{array}{ll}
    A^{T}A - C^{T}C = \mbox{\bf 1},&
    A^{T}B - C^{T}D = \mbox{\bf 0},\\
    B^{T}A - D^{T}C = \mbox{\bf 0},&
    B^{T}B - D^{T}D= -\mbox{\bf 1} \\
    \end{array}
\eq
These relations imply that the blocks of $t_{i}$ satisfy
\be   \begin{array}{ll}
    a^{+} a - c^{+}c = \mbox{\bf 1},&
    a^{+}b - c^{+}d = \mbox{\bf 0},\\
    b^{+}a - d^{+}c= \mbox{\bf 0},&
    b^{+}b - d^{+}d= -\mbox{\bf 1}\\
     \end{array}
\eq
This means that $t_{i}$ is pseudo--unitary, $t_{i} \in$
$U(p_{i},\; q_{i})$. Therefore the group of those deformations
in the $i$--th eigenspace that commutes with the twist
is (at least locally) isomorphic to $U(p_{i},\; q_{i})$.

Combining the above results for all
the blocks in the decomposition of a general
$\cal T$ $\in$ $O(16 +d,\; d)$ we have shown that those deformations $\cal T$
commuting with the twist $\Theta$
form a subgroup isomorphic to
\be
  \bigotimes_{i=1}^{K} U(p_{i},\; q_{i}) \otimes
   O(d_{-1}^{(L)},\;d_{-1}^{(R)}) \otimes
   O(d_{1}^{(L)},\;d_{1}^{(R)})
\eq
where $K$ is the total number of distinct pairs of complex
eigenvalues.

However, those deformations $\cal T$
which are pure rotations, $\cal T$ $\in$ $O(16+d) \otimes O(d)$,
do not
change the physical content of a model. To get the (untwisted)
moduli space (up to modular transformations) we have to
factorize this subgroup. The resulting coset space is given by
\be
{\cal M}_{O}(\Theta) \simeq
\bigotimes_{i=1}^{K}
\frac{ SU(p_{i},\; q_{i}) }{SU(p_{i}) \otimes SU( q_{i}) \otimes U(1)}
\otimes
\frac{SO(d_{- 1}^{(L)},\;d_{- 1}^{(R)})}
{SO( d_{- 1}^{(L)} ) \otimes SO( d_{- 1}^{(R)} )} \otimes
\frac{SO(d_{1}^{(L)},\;d_{1}^{(R)})}
{SO( d_{1}^{(L)} ) \otimes SO( d_{1}^{(R)} )}.
\label{ModuliSpaceO}
\eq
Note that we have made use of
the local isomorphisms $U(p,\;q)$ $\simeq$
$SU(p,\;q) \otimes U(1)$ and $O(p,\;q) \simeq SO(p,\;q)$ to
bring our result into the form usually used in the supergravity
literature.
As claimed above, the local structure
of the untwisted moduli space is completely determined by the
eigenvalues of the twist $\Theta$.
The dimension of the moduli space is
\be   \mbox{dim} ( {\cal M}_{O}(\Theta) ) =
       2 \sum_{i=1}^{K} p_{i} q_{i} + d_{-1}^{(L)} d_{-1}^{(R)}
      + d_{1}^{(L)} d_{1}^{(R)}
\label{DimMO}
\eq
It only depends on the multiplicities of the eigenvalues of
$\Theta$. Moduli do only exist if an eigenvalue appears both
in the left and in the right part of the twist.

We can now compare our result (\ref{ModuliSpaceO})
with the coset spaces found
in \cite{FKP,CveLouOvr}
for the
${\bf Z}_{N}$ orbifold compactifications without Wilson lines
yielding ${\cal N} = 1$ and ${\cal N} = 2$ space-time supersymmetry.
In order to
do so, we simply have to restrict ourselves
to an $O(d,d)$ subsector and to set $d=6$ as well as
$\Theta_{L} = \Theta_{R}$
$=\theta$.  Then, by
plugging into (\ref{ModuliSpaceO})
the wellknown eigenvalues of the symmetric
${\bf Z}_{N}$ twists leading to ${\cal N} =1,2$ space-time supersymmetry
\cite{DHVW2,ErlKle},
we find that all the results agree
\footnote{In fact, it is easily seen that the
form of the world sheet action which is crucial in the approach
of \cite{CveLouOvr} depends only on the eigenvalues
of $\theta$ and their multiplicities.},as expected.
As a straightforward generalization we can now similarly
write down the cosets for all these models with
continuous Wilson lines turned on.
The result will, of
course, now also depend on the gauge twist $\theta'$ and its
eigenvalues. Since the choice of a gauge twist is
also constrained by world sheet modular invariance one has
to proceed as follows.
First, one has to find all $E_{8} \otimes E_{8}$  Weyl twists
$\theta'$ which
have the required order and lead to modular invariant
twists $\Theta$. Then one has to calculate their eigenvalues in order
to get the coset.
To carry out this program will require some work because
there are, in general, a lot of gauge twists satisfying the
constraints from modular invariance
, especially for higher $N$. Based on \cite{KKKOT}
one can estimate that there will be roughly 500 models.
This will, therefore, be the subject of a later publication
\cite{CarLuMoh}.
However, to give an explicit example, we will list the
cosets for all modular invariant ${\bf Z}_{3}$
orbifold compactifications with ${\cal N}=1$ space-time
supersymmetry.
This is easy to do, since both $\theta$ and $\theta'$ consist
of several copies of the $A_{2}$ coxeter twist\footnote{
$A_{2}$ is the complex form of $su(3)$.}.
This is a rotation by 120 degrees and therefore has the
eigenvalues $\exp(\pm 2 \pi i/3)$. More precisely, $\theta$
contains three copies of this twist and the gauge twist
$\theta'$ is constrained by modular invariance to contain
0, 3 or 6 further copies. This leads to five inequivalent
models \cite{IbaNilQue2}.
In table (\ref{Z3}) we list these twists together
with the corresponding moduli spaces and the maximal and
minimal gauge group. The maximal gauge group is the one
of the model with vanishing Wilson lines and can be
found in \cite{IbaNilQue2} or in the table of
$E_{8}$ automorphisms given in \cite{HolMyh}.
The minimal gauge group is the
one for generic (purely continuous) Wilson lines
and can be calculated using the method introduced in \cite{Moh2}.

\begin{table}
\[
\begin{array}{|c|c|c|c|} \hline
\mbox{Gauge Twist} & \mbox{Coset}  & \mbox{Max. Gauge Group} &
\mbox{Min. Gauge Group} \\ \hline \hline
\emptyset \otimes \emptyset &
\frac{ SU(3,\;3)}{ SU(3) \otimes SU(3) \otimes U(1) } &
E_{8} \otimes E_{8}\,' &
E_{8} \otimes E_{8}\,' \\ \hline
A_{2}^{2} \otimes A_{2} &
\frac{ SU(6,\;3)}{ SU(6) \otimes SU(3) \otimes U(1) } &
(SO(14) \otimes U(1) ) \otimes (E_{7} \otimes U(1))\,' &
(SU(3) \otimes SU(3) ) \otimes E_{6}\,' \\ \hline
A_{2}^{3} \otimes \emptyset &
\frac{ SU(6,\;3)}{ SU(6) \otimes SU(3) \otimes U(1) } &
(E_{6} \otimes SU(3)) \otimes E_{8}\,' &
SU(3) \otimes E_{8}\,' \\ \hline
A_{2}^{3} \otimes A_{2}^{3} &
\frac{ SU(9,\;3)}{ SU(9) \otimes SU(3) \otimes U(1) } &
(E_{6} \otimes SU(3)) \otimes (E_{6} \otimes SU(3))\,' &
SU(3) \otimes SU(3)\,' \\ \hline
A_{2}^{4} \otimes A_{2}^{2} &
\frac{ SU(9,\;3)}{ SU(9) \otimes SU(3) \otimes U(1) } &
SU(9) \otimes (SO(14) \otimes U(1) )\,' &
(SU(3)\otimes SU(3))\,' \\ \hline
\end{array}
\]
\caption{Table of all ${\bf Z}_{3}$ orbifold compactifications
with ${\cal N} =1$ space-time supersymmetry.}
\label{Z3}
\end{table}

Let us conclude with two further remarks. The first is
that the formula (\ref{DimMO}) coincides with the one derived
in \cite{Moh2} for orbifold compactifications with continuous
Wilson lines and a special choice of the gauge twist.
Conversely the results of this section indicate that it
should be possible to generalize the results of \cite{Moh2}
to the general asymmetric case.

The final remark concerns the K\"{a}hlerian structure
of the cosets given in
(\ref{ModuliSpaceO}).
Whereas the $SU$ cosets
are K\"ahlerian for any value of the parameters, the $SO$
cosets are K\"ahlerian if they are isomorphic to $SU$ cosets
by some accidental isomorphism of (low dimensional) Lie groups
or if one of the
parameters equals 2, that is for cosets \cite{Gil}
\be     \frac{SO(p,\; 2)}{SO(p) \otimes SO(2)}
\eq
For ${\cal N}=1$ supersymmetric orbifold compactifications
it is well known that the eigenvalue $+1$ does not appear,
whereas $-1$ does only appear with multiplicities 0 or 2
\cite{DHVW2,ErlKle}.
Therefore, the moduli spaces
of ${\cal N}=1$ ${\bf Z}_{N}$ orbifold
compactifications with continuous
Wilson lines  found here are all K\"ahlerian. For general asymmetric
${\bf Z}_{N}$ orbifolds the situation is less clear, since
the necessary and sufficent condition for ${\cal N} =1$
space-time supersymmetry
is not known. Our result suggests that the only real
eigenvalue $\Theta_{L}$ may have is $-1$, with multiplicity
0 or 2 as in the compactification case.
Note, however, that our investigations have been restricted to
the untwisted sector and that it has been pointed out
recently \cite{Sas} that space-time supercharges may also appear
from twisted sectors in asymmetric orbifolds.

\setcounter{equation}{0}

\section{Modular symmetries}

\hspace*{.3in}
The fact that the naive moduli spaces will contain several copies of
the same model is clear from the beginning since one can easily imagine
that there will be large deformations ${\cal T} \in O(16+d,\,d)$ which are
at the same time automorphisms of the Narain lattice and therefore
do not lead to a different model. Therefore all transformations
of the type
\be   {\cal T} \in O(16+d,\,d) \cap \mbox{AUT}(\Gamma)
\eq
are symmetries of the toroidal moduli space ${\cal M}_{T}$.
Those which also fulfill the normalizer constraint for a twist $\Theta$
\be  {\cal T}
\in {\cal N}(\la \Theta \ra, O(16+d,\;d) \cap \mbox{AUT}(\Gamma))
\eq
are then symmetries of the orbifold moduli space
${\cal M}_{O}(\Theta)$.

In the following we will recall and extend the analysis performed by
Spalinski \cite{Spa}. In this approach one first finds the action of modular
transformations on the quantum numbers and then derives the induced
action on the moduli themselves.
To do so, one first writes down the indefinite bilinear form
in the lattice basis (in matrix notation) as
\be {\bf P}_{L}^{2} - {\bf P}_{R}^{2} =
    v^{T} H v
\eq
Here $v$ is a vector consisting of the $16 + 2d$ integer quantum
numbers which label a state,
\be    v^{T}  =  (  q^{A}, n^{i}, m_{i} ) \in {\bf Z}^{16 + 2d}, \;\;\;
       A= 1,\ldots, 16, \; i = 1,\ldots, d,
\eq
and $H$ is the lorentzian lattice metric of $\Gamma$ given as
\be H  = \left( \begin{array}{ccc}
          (l_{A}, l_{B}) & (l_{A}, \ov{k}_{j}) & (l_{A}, k^{n}) \\
          (\ov{k}_{i}, l_{B})&  (\ov{k}_{i}, \ov{k}_{j}) &
                                  (\ov{k}_{m}, k^{n}) \\
          (k^{m}, l_{B}) &  (k^{m}, \ov{k}_{j})&
                     (k^{m}, k^{n}) \\
          \end{array} \right)
         =  \left( \begin{array}{ccc}
           C & 0 & 0 \\ 0 & 0 & I \\ 0 & I & 0 \\
           \end{array} \right)
\label{M5}
\eq
where $C$ is the Cartan matrix of $E_{8} \otimes E_{8}$ and I
is the $d \times d$ unit matrix.
Modular symmetry transformations $\cal T$ can now also be described
in terms of their matrices $\Omega$ with respect to the lattice basis,
which act as
\be v \rightarrow v' =  \Omega^{-1} v
\eq
on the quantum numbers. To be a symmetry, the matrix $\Omega$ must be
integer valued,
\be   \Omega \in GL(16 + 2d, {\bf Z})  \Leftrightarrow {\cal T} \in
         \mbox{AUT}(\Gamma)
\eq
and it must leave the indefinite bilinear form invariant
\be   \Omega^{T} H \Omega  = H  \Leftrightarrow {\cal T} \in
      O(16 +d,d)
\label{M8}
\eq
These two conditions combined define a matrix group,
\be   G_{T} = \{ \Omega \in GL(16 + 2d,{\bf Z}) | \Omega^{T} H \Omega
                   = H \}
\label{M9}
\eq
which
is called the modular invariance group of toroidal compactifications.
It is usually denoted by $O(16 + d,d; {\bf Z})$ for obvious reasons,
although it is not a group of pseudo--orthogonal matrices.
To get the modular invariance group for an orbifold one simply has
to add the normalizer constraint.  Then
\be  G_{O} = {\cal N}( \la \Theta \ra, G_{T} )
    = \{ \Omega \in GL(16 +2d, {\bf Z}) | \Omega^{T} H \Omega = H,
\;\;\;      \Omega^{-1} \Theta \Omega = \Theta^{k}, \exists k:\;
          1 \leq k < N \}
\label{M10}
\eq
This last formula was given by Spalinski \cite{Spa}
but without mentioning its
group theoretical interpretation.
The calculation of such normalizers is in general a difficult task
which must be done case by case.  Nevertheless,
several examples have been
discussed in the literature
\cite{FerFreSor,Spa,ErlSpa,Erl,BLST1,BLST2} for models with no or
with discrete Wilson lines.
As pointed out in \cite{ErlSpa}
a factorization of the modular invariance group
into factors corresponding to different
eigenvalues of the twist can only be expected if the underlying
lattice itself decomposes into an orthogonal direct sum, which is
not the case generically. Therefore the local decomposition of
orbifold moduli spaces into a product of coset spaces does not
imply a corresponding decomposition of $G_{O}$.

The second step is to deduce the action of modular symmetry
transformations on the moduli. This can be done through the mass
formula (\ref{m15}). Therefore we express the euclidean bilinear form
in lattice coordinates as
\be  {\bf P}_{L}^{2} + {\bf P}_{R}^{2} = v^{T} \Sigma v
\label{M11}
\eq
The mass matrix $\Sigma$ which encodes the complete moduli dependence
of the whole spectrum is the euclidean lattice metric
\be \Sigma   = \left( \begin{array}{ccc}
          l_{A} \cdot l_{B} & l_{A} \cdot \ov{k}_{j} & l_{A} \cdot k^{n} \\
          \ov{k}_{i} \cdot l_{B} & \ov{k}_{i} \cdot \ov{k}_{j} &
                                  \ov{k}_{m} \cdot k^{n} \\
          k^{m} \cdot l_{B} &  k^{m} \cdot \ov{k}_{j}&
                     k^{m} \cdot k^{n} \\
          \end{array} \right)
\label{M12}
\eq
Introducing matrix notation and working out the scalar products one
gets
\be  \Sigma = \left( \begin{array}{ccc}
              C + \frac{1}{2} A G^{-1} A^{T} &
              - \frac{1}{2} A G^{-1} D^{T} &
              -  \frac{1}{2} A G^{-1} \\
              - \frac{1}{2} D G^{-1} A^{T} &
               \frac{1}{2} D G^{-1} D^{T} &
              I + \frac{1}{2} D G^{-1} \\
              - \frac{1}{2} G^{-1} A^{T} &
              I + \frac{1}{2} G^{-1} D^{T} &
              \frac{1}{2} G^{-1} \\
              \end{array} \right)
\label{Sigma}
\label{M13}
\eq
Here $G = (G_{ij})$ and $G^{-1} = (G^{ij})$ are the lattice
metrics\footnote{Note that some authors
choose the lattice metric of $\Lambda$ to be 2$G$. With this
convention, some of the matrix elements differ by a factor 2.}
of the compactification lattice $\Lambda$ and its dual $\Lambda^{*}$.
The matrices $A=(A_{Ai})$ and $D = (D_{ij})$ were defined in
(\ref{DefA}) and (\ref{DefD}), respectively.

The action of a modular symmetry transformation on the euclidean
bilinear form is given by
\be v^{T} \Sigma v \rightarrow v^{T} \Omega^{T,-1} \Sigma'
           \Omega^{-1} v
         \stackrel{!}{=} v^{T} \Sigma v
\label{M14}
\eq
Note that the moduli dependent matrix $\Sigma$ will in general
also transform, if the deformation $\Omega$
is not a pure rotation of
the Narain lattice. The fact that the deformations we consider
are symmetry transformations and therefore must leave the
euclidean bilinear form invariant fixes the transformation
law of $\Sigma$
to be
\be  \Sigma \rightarrow \Sigma' = \Omega^{T} \Sigma \Omega
\label{M15}
\eq
with $\Omega \in G_{T}, G_{O}$ respectively \cite{Spa}.
Since the functional dependence of $\Sigma$ on the moduli
is known via (\ref{Sigma}),
this allows one, in principle, to calculate the transformation law
of the moduli. But as the dependence is quite complicated and
nonlinear, this is tedious to do in practice. Since $\Sigma$ is
symmetric it is tempting to try to factorize it in the form
\beq
\Sigma = \phi^{T} \phi
\label{Fac}
\eeq
hoping that the moduli dependence of
$\phi$ might be simpler. In the case of bosonic strings the construction
given by Giveon, Porrati and Rabinovici \cite{GPR} does precisely this.
But in order to apply their bosonic result
to heterotic strings one has do embed the heterotic string into the
bosonic one and calculations are still complicated.

We will, therefore, use a different approach where one directly works
with the heterotic string. It is also motivated by the question
how the moduli $G_{ij}, B_{ij}, {\bf A}_{i}$
are related to standard (homogenous and
projective) coordinates on cosets, which are known to have a
simple transformation under the group action.
To explain why these two questions are closely related
let us write down the euclidean bilinear form in matrix notation,
but now with respect to an orthonormal frame, and then apply a
deformation (not a modular transformation)
${\cal T} \in O(16+d,d;{\bf R})$.  Then
\be {\bf P}^2_L + {\bf P}^2_R =
u^{T} u \rightarrow u^{T} {\cal T}^{T} {\cal T} u
\label{DeformONB}
\eq
Note that we have expressed the deformed bilinear form in terms of the
old coordinates $u$. The moduli dependence is now completely given by
the symmetric matrix ${\cal T}^{T} {\cal T}$. The same deformation
can be described
with respect to lattice coordinates $v$:
\be   {\bf P}_{L}^{2} + {\bf P}_{R}^{2} = v^{T} \Sigma_{0} v
      \rightarrow v^{T} \Sigma v
\label{DeformLB}
\eq
We consider $\Sigma_{0}$ as a fixed reference background and $\Sigma$
as a function of the moduli which describe the continuous
deformations of $\Sigma_{0}$. If $N$ is the basis transformation
connecting the $u$ and the $v$ frame by $u = Nv$ then, by combining
(\ref{DeformONB}) and (\ref{DeformLB}),
we get a decomposition of $\Sigma$ which has
the desired form (\ref{Fac})
\be   \Sigma = N^{T} {\cal T}^{T} {\cal T} N
\label{NT}
\eq
As is well known from
the case of the Lorentz group, elements of pseudo orthogonal groups
can be decomposed into a rotation $R  \in O(16+d) \otimes O(d)$
and a "boost" $B$.  The latter can be used as a coset representative.
In the case of (\ref{NT})
the rotational part always cancels out
\be {\cal T}= R B \Rightarrow {\cal T}^{T} {\cal T} = B^{T} R^{T} R B =
    B^{T} B
\eq
which again shows that the spectrum only depends on the coset.
We can therefore expect that it is possible to factorize $\Sigma$
in terms of a matrix $\phi$ which is a product of a coset
representative $B \simeq R B$ and a basis transformation $N$.

While this consideration has told us what $\phi$ should be, it didn't
say how to construct it. There is however one obvious way to factorize
$\Sigma$ as in (\ref{Fac}). Namely, we introduce an orthonormal basis
\be  \wh{\bf e}_{a} = \left( {\bf e}_{a}, {\bf 0}; {\bf 0} \right),
      \;\;
     \wh{\bf e}_{\mu}^{(L)} = \left( {\bf 0}, {\bf e}_{\mu}; {\bf 0}
                              \right), \;\;
     \wh{\bf e}_{\mu}^{(R)} = \left( {\bf 0}, {\bf 0}; {\bf e}_{\mu}
                              \right)
\label{Orthbas}
\eq
($a = 1, \ldots, 16$, $\mu = 1,\dots,d$),
and expand all the vectors appearing in (\ref{M12}) in terms of it.
This yields:
\be   \Sigma =
      \left( \begin{array}{ccc}
      l_{A} \cdot {\bf e}_{a}   &
      l_{A} \cdot {\bf e}_{\mu}^{(L)} &
      l_{A} \cdot {\bf e}_{\nu}^{(R)} \\
      \ov{k}_{i} \cdot {\bf e}_{a}   &
      \ov{k}_{i}\cdot {\bf e}_{\mu}^{(L)} &
      \ov{k}_{i}\cdot {\bf e}_{\nu}^{(R)} \\
      k^{m} \cdot {\bf e}_{a}   &
      k^{m}\cdot {\bf e}_{\mu}^{(L)} &
      k^{m}\cdot {\bf e}_{\nu}^{(R)} \\
      \end{array} \right)
      \left( \begin{array}{ccc}
      {\bf e}_{a}\cdot l_{B}   &
      {\bf e}_{a}\cdot \ov{k}_{j}   &
      {\bf e}_{a}\cdot k^{n}  \\
      {\bf e}_{\mu}^{(L)} \cdot l_{B}   &
      {\bf e}_{\mu}^{(L)} \cdot \ov{k}_{j}   &
      {\bf e}_{\mu}^{(L)} \cdot k^{n}  \\
      {\bf e}_{\nu}^{(R)} \cdot l_{B}   &
      {\bf e}_{\nu}^{(R)} \cdot \ov{k}_{j}   &
      {\bf e}_{\nu}^{(R)} \cdot k^{n}  \\
      \end{array} \right)
      = \phi^{T}\phi
\eq
Working out the scalar products we get
\be \phi^{T} = \left( \begin{array}{ccc}
     {\cal E} & -\frac{1}{2} A E^{*} & - \frac{1}{2} A E^{*} \\
     A^T {\cal E}^{*} & (2 G + \frac{1}{2} D) E^{*} & \frac{1}{2} D E^{*} \\
     0 & \frac{1}{2} E^{*} & \frac{1}{2} E^{*} \\
     \end{array} \right)
\label{cosetrep}
\eq
The ambiguity in the factorization of $\Sigma$ is reflected by
the appearence of the $n$--bein matrices
${\cal E} = ({\bf e}_{A} \cdot {\bf e}_{a})$,
$E = ({\bf e}_{i} \cdot {\bf e}_{\mu})$ and their duals
${\cal E}^{*}={\cal E}^{T, -1}$, $E^{*} = E^{T,-1}$.
Since we expect $\phi^{T}$ to be a coset representative we now
compare (\ref{cosetrep}) to the well known standard form of such an
object \cite{Gil}.
Although (\ref{cosetrep}) is not in the standard form
of a coset representative, its
structure is similar enough to allow for the
construction of analogues of
standard homogenous and projective coordinates.
Let us first introduce
\be X = \left( \begin{array}{c}
        - \frac{1}{2} A E^{*} \\ \frac{1}{2} D E^{*} \\
         \end{array}  \right),\;\; \;\;\;
    Y = \left( \frac{1}{2} E^{*} \right)
\eq
Note that all the moduli appear in  $X$, so that the whole
information is encoded in it. But $X$ is not a nice coordinate.
However, following the standard procedure described in \cite{Gil},
the $(16 + 2d) \times d$ matrix
\be \left( \begin{array}{c}
     X \\ Y \\
    \end{array} \right) =
    \left( \begin{array}{c}
      X_1 \\
      X_2 \\
       Y \\
     \end{array} \right)=
    \left( \begin{array}{c}
     - \frac{1}{2} A E^{*} \\
     \frac{1}{2} D E^{*} \\
     \frac{1}{2} E^{*} \\
     \end{array} \right)
\label{M22}
\eq
which consists of the last $d$ columns of $\phi^{T}$,
should be a homogenous coordinate. This means that it must
transform linearly under the left action of the group modulo
rotations acting from the right.
In our case the group action is given by
\be   \phi^{T} \rightarrow \Omega^{T} \phi^{T}
\eq
where
\be  \Omega \in \{ M \in GL(16 + 2d; {\bf R}) |
              M^{T} H M = H \} \simeq O(16 + d, d; {\bf R})
\eq
Decomposing $\Omega^{T}$ into appropriate blocks
\be \Omega^{T} = \left( \begin{array}{cc}
            a & b \\ c & d \\
            \end{array} \right)
\label{M25}
\eq
we see that our coordinate transforms indeed linearly as
\be  \left( \begin{array}{c}
     X \\ Y \\ \end{array} \right)
     \rightarrow
     \left( \begin{array}{c}
     a X + b Y \\ c X + d Y \\
     \end{array} \right)
\eq
This linear transformation law is (as usual for homogenous
coordinates) achieved by treating $X$ and $Y$ as being
independent of each other.  It can be checked, however, that
$X$ and $Y$ are really constrained
to satisfy
\be   X_1^T C^{-1} X_1 + X_2^T Y +
     Y^T X_2 = - I
\eq
where $C^{-1}$ is the inverse of the Cartan matrix
of $E_{8} \otimes E_{8}$.
These are, in fact, the constraint equations for a
$\frac{ O(16 + d, d) }{O(16 + d) \otimes O(d)}$
coset as is well known from both the mathematical \cite{Gil} and
the supergravity literature \cite{PorZwi}.
Such equations are crucial for
the construction of supergravity actions and especially of the associated
K\"ahler potential. As an example, we will in section 5 discuss the case
of an $\S4$ coset in big detail.

 From the homogenous coordinate it is easy to construct a projective
one, that is a coordinate that transforms under the group action
by fractional linear transformations \cite{Gil}.
Defining
\be  Z = X Y^{-1} = \left( \begin{array}{c}
     - A \\ D
     \end{array} \right)
\label{Pro}
\eq
we see that
the group acts in fact on $Z$ by fractional linear transformations
\be Z \rightarrow (aZ + b) (cZ + d)^{-1}
\label{Tra}
\eq
Note also that the dependence on rotations from the right
that the homogenous coordinate still had,
has completely cancelled out in (\ref{Pro}).
This is manifest since the
$d$--bein variable $E^{*}$ has disappeared.
This is the second typical feature of a projective
coset coordinate \cite{Gil}.

One useful application of the projective coordinate is that
the transformation properties of the moduli can be deduced very
simply from it. To pass from the deformation group $O(16 + d,\, d)$
to the modular invariance group $O(16 + d,\,d; {\bf Z})$
one simply has to restrict to the subgroup of
integer valued matrices. As an example let us calculate
the transformation of the moduli under the duality transformation
which generalizes the well known $R \rightarrow \frac{1}{2R}$
duality known from compactification on the circle.
On the level of quantum numbers the generalized duality transformation
exchanges winding and momentum numbers while leaving the charges
invariant. The corresponding matrix in $O(16 + d,d;{\bf Z})$
is obviously
\be  \Omega^{T}  =\left( \begin{array}{c|c}
     a & b \\ \hline c & d\\
     \end{array} \right) =
     \left( \begin{array}{cc|c}
      I & 0 & 0 \\
      0 & 0 & I \\ \hline
      0 & I & 0 \\
      \end{array} \right)
\eq

By acting with $\Omega^{T}$ on $Z$ as in (\ref{Tra})
one immediately  gets that
\be      Z = \left( \begin{array}{c}
         -A \\ D
          \end{array} \right)
         \rightarrow
          \left( \begin{array}{c}
          - A D^{-1} \\ D^{-1} \\
          \end{array} \right)
\label{M31}
\eq
a result which is much harder to derive by other methods.

In the orbifold case the transformation law of the real moduli
remains, of course, the same but one has to check that the
matrix acting on the quantum numbers fulfills the normalizer
condition (\ref{m13}). There is, however, another problem (for models with
${\cal N} = 1$ supersymmetry)
since one would
like to use a complex parametrization of the moduli. This will be
discussed in the next section.

As a second application of the projective coordinate (\ref{Pro})
let us rederive the global structure of the toroidal moduli space.
It is well known that the action of the modular invariance group
on the moduli is not faithful \cite{Spa}. Therefore the true modular symmetry
group ${\cal G}_{T}$ is a factor group of $G_{T}$ by those
transformations which act trivially on the toroidal moduli
. It is known that \cite{GPR}
\be {\cal G}_{T} = PO(16 + d,d; {\bf Z}) =
    O(16 + d,d; {\bf Z}) / \{ \pm I   \}
\eq
Whereas it is obvious that $-I$ acts trivially on the moduli,
it is less obvious that there are no further trivial transformations.
However, by using the projective coordinate $Z$ it becomes clear that,
if one requires
\be (aZ + b) (cZ + d)^{-1} = Z
\eq
for all $Z$,  then it follows that
$b=0$, $c=0$ and that either $a=I$, $d=I$ or
$a=-I$, $d=-I$.

In the orbifold case the connection between $G_{O}$ and
${\cal G}_{O}$ is an open problem \cite{Spa}.
Clearly the group generated
by the twist $\Theta$ acts trivially, by construction, but
this is all one knows in general. Of course, one is
ultimately interested in ${\cal G}_{O}$ because it is this group
which acts on the moduli and, therefore, decides both about the global
geometry of moduli space and the form of effective supergravity
actions. At least in the cases where $G_{O}$ factorizes into
groups acting on the cosets in the decomposition
(\ref{ModuliSpaceO})
it should be
possible to make a general statement about the connection
between  $G_{O}$ and ${\cal G}_{O}$, but we will not try to do
so in this paper.   We will, in the following,
 rather focus on (\ref{ModuliSpaceO}) and
deal with the problem of its complexification
and the derivation of the associated K\"ahler potentials.

\setcounter{equation}{0}

\section{K\"{a}hler potentials for $Z_N$ orbifold compactifications
with continuous Wilson lines}

\hspace*{.3in} As we have seen above, the untwisted moduli space of
$Z_N$ orbifold compactifications with continuous Wilson lines preserving
${\cal N}=1$ space-time supersymmetry
is locally
given by a direct product of $\SUm$ and $\SOd$ cosets.  Each
of these cosets is a  K\"{a}hlerian manifold,  that is,
its metric $g_{\phi \bar{\phi}}$ is locally expressible as
$g_{\phi \bar{\phi}} = \partial_{\phi} \partial_{\bar{\phi}}
K({\phi},{\bar{\phi}})$ in some complex coordinate system $(
{\phi}, {\bar{\phi}})$.  $K$ denotes its K\"{a}hler potential.
The
K\"{a}hler structure of the untwisted
moduli space is then determined by the
full K\"{a}hler potential given by the
sum of the K\"{a}hler potentials of the individual cosets.
The
K\"{a}hler potential, on the other hand,
is also one of the three fundamental functions which
describe the tree-level couplings of generic matter multiplets
to $4D,{\cal N}=1$ supergravity, as is well known
\cite{CreJul,CreFer}. Thus, in order
to determine the tree-level low-energy Lagrangian describing the coupling
of the $Z_N$ orbifold
moduli fields to supergravity, knowledge of the associated
K\"{a}hler potential is  crucial.  This section is devoted towards
explicitly constructing the K\"{a}hler potentials for some of the moduli
spaces discussed in the previous section.  Namely, we will focus on
the $Z_N$ orbifold compactifications
for which the internal 6-torus $T_6$ can be decomposed
into a direct sum $T_4 \oplus T_2$.  All such cases are given in table
(\ref{T6})
\cite{ErlKle}.  Then, by using well known techniques
\cite{PorZwi,FerPor,FerKouLuZw}, we will derive
the K\"{a}hler potentials for the moduli spaces associated with the
underlying 2-torus $T_2$.
We will first
focus on the $\SOd$ cosets.  As stated earlier, they arise
when the twist operating on $T_2$ has two
eigenvalues of
$-1$.  It is well known \cite{GPR}
that, in the case when no continuous Wilson
lines are turned on, the resulting $\SO2$ coset factorises into
$\SO2 = \left( {\SU} \right)_T \otimes \left( {\SU}\right)_U$,
each of them being coordinatised by one
complex modulus, $T$ and $U$, respectively.  The associated K\"{a}hler
potential is then simply given by the sum of two individual
 K\"{a}hler potentials.
When turning on Wilson
lines, however,
the resulting coset $\SOd$ will in general not factorise anymore, and
the resulting K\"{a}hler potential will be much more complicated.
Nevertheless, one still expects to find one modified $T$ and
one modified $U$ modulus among the complex coordinates of the $\SOd$
coset.  This is in fact the case, as we shall see below.
We will, for concreteness,
discuss the $\S4$ and the $\P3$
cosets in great detail. They are the simplest non-trivial
ones occuring when turning on the two Wilson lines ${\bf A}_i$
associated with the two directions of the underlying 2-torus $T_2$.
Any other $\SOd$ coset, however, can in
principle be analysed along very similar lines, although arriving at explicit
results for the K\"{a}hler potential might be quite tedious.
Finally, at the end of this section, we will discuss the K\"{a}hler
potential for the $\SUS$ cosets
with Wilson lines turned on. They occur whenever the twist acting on the
underlying 2-torus $T_2$ doesn't have eigenvalues $-1$.
We will, for concreteness, discuss the $\SK$ coset in great detail.
Its discussion, however, can be generalised to any other $\SUS$
coset in a
straightforward manner.

\begin{table}
\[
\begin{array}{|c|c|c|} \hline
\mbox{Case} & \mbox{Twist}  & \mbox{Lattice} \\ \hline \hline
 2 \;\;\;\;\; Z_4 & (Z_2^{(1)},Z_2^{(1)},Z_4^{(2)},Z_4^{(2)}) &
A_1 \times A_1 \times B_2 \times B_2\\ \hline
 5 \;\;\;\;\; Z_6 & (Z_3^{(2)},Z_6^{(2)},Z_6^{(2)})&
A_2  \times G_2 \times G_2 \\ \hline
 7 \;\;\;\;\; Z_6' & (Z_2^{(1)},Z_2^{(1)},Z_3^{(2)},Z_6^{(2)}) &
A_1 \times A_1 \times A_2 \times G_2 \\ \hline
12 \;\;\;\;\; Z_8 & (Z_4^{(2)},Z_8^{(4)})&
B_2 \times B_4\\ \hline
14 \;\;\;\;\; Z_8' & (Z_2^{(1)},Z_2^{(1)},Z_8^{(4)}) &
A_1 \times A_1 \times B_4 \\ \hline
16 \;\;\;\;\; Z_{12} & (Z_3^{(2)},Z_{12}^{(4)}) & A_2 \times F_4 \\ \hline
18 \;\;\;\;\; Z_{12}' & (Z_2^{(1)},Z_2^{(1)},Z_{12}^{(4)}) & A_1 \times
A_1 \times F_4 \\ \hline
\end{array}
\]
\caption{$Z_N$ orbifold compactifications
with ${\cal N} =1$ space-time supersymmetry and $T_6=T_2 \oplus T_4$,
as given
in the classification of
[19].}
\label{T6}
\end{table}

Let us begin by explicitly constructing a suitable
set of complex coordinates for a $\SOd$ coset along the lines of
\cite{PorZwi,FerPor,FerKouLuZw}.
A set of real coordinates $\Upsilon_{\mu}\,^I$ for a general coset space
$\SOd$ was found in (\ref{M22}) and is given by
\beqa
 (\Upsilon_{\mu}\,^I) =
\left( \begin{array}{c}
        {-\cal{L}}_{\mu a} \\
     - (L + \tilde{L})_{\mu}\,^i  \\
     \frac{1}{2} (L - \tilde{L})_{\mu}\,^i \\
      \end{array}  \right) =
 \left( \begin{array}{c}
         - \frac{1}{2} {\cal{A}}_{ak} E_{\mu}\,^k \\
           \frac{1}{2} D_{ij} E_{\mu}\,^j \\
          \frac{1}{2} E_{\mu}\,^i    \end{array}  \right)
\label{K1}
\eeqa
The real moduli
matrix $D_{ij}$, associated with the underlying 2-torus $T_2$,
is given in (\ref{DefD}).  $E^{*}=({\bf e}_{\mu} \cdot
{\bf e}^k)= (E_{\mu}\,^k)$ denotes
the inverse zweibein.  Note, too, that we have, for later convenience,
introduced component Wilson lines ${\cal A}_{ai}$ as
\beq
{\cal A}_{ai}= {\bf e}_a \cdot {\bf A}_i , \;\;\;\; a=1,...,16
\label{K2}
\eeq
They are defined relative to the orthonormal basis ${\bf e}_a$ introduced in
(\ref{Orthbas}).
The ${\cal A}_{ai}$ are not to be confused with the $A_{Ai}$
introduced in (\ref{DefA}).  The latter
ones are defined relative to the lattice basis ${\bf e}_A$.
Note that (\ref{AiAj})
can be equivalently expressed as
\beq
{\bf A}_i \cdot {\bf A}_j = {\cal A}^a\,_i {\cal A}_{aj}
\eeq
Also note that $r$ can take any value between $2 \leq r \leq 18$.

It can be readily checked that the
$\Upsilon_{\mu}\,^I$ satisfy the relation
\beq {\theta}_{IJ} \Upsilon_{\mu}\,^I
\Upsilon_{\nu}\,^J = - {\delta}_{\mu \nu}
\label{K3}
\eeq
where the metric tensor ${\theta}_{IJ}$ for pulling down indices
is given by
\beqa
 ({\theta}_{IJ})=
  \left( \begin{array}{ccc}
  {\delta}^{ab} & 0 & 0  \\
  0 & 0 &{\delta}^j \,_i \\
  0 & \delta_i \,^j & 0   \end{array}  \right)
\label{K4}
\eeqa
Disentangling the coordinates
$L$ and $\tilde{L}$ \cite{PorZwi}
yields a new set of real coordinates $\Upsilon_{\mu}\,^I$
given by
\beqa
(\Upsilon_{\mu}\,^I) =
\left( \begin{array}{c}
        {\cal{L}}_{\mu a} \\
       {\tilde{L}}_{\mu}\,^i  \\
       L_{\mu}\,^i  \\
      \end{array}  \right) =
 \left( \begin{array}{c}
          \frac{1}{2} {\cal{A}}_{ak} E_{\mu}\,^k \\
           \frac{1}{2}(-E_{\mu}\,^i + E_{\mu i} + E_{\mu}\,^j B_{ji}
+ \frac{1}{4} E_{\mu}\,^j ({\A}^a\,_j {\A}_{ai})) \\
          \frac{1}{2}(E_{\mu}\,^i + E_{\mu i} + E_{\mu}\,^j B_{ji}
+ \frac{1}{4} E_{\mu}\,^j ({\A}^a\,_j {\A}_{ai}))    \end{array}  \right)
\label{K5}
\eeqa
and satisfying the $\SOd$ coset relation
\beq {\theta}_{IJ} \Upsilon_{\mu}\,^I
\Upsilon_{\nu}\,^J = - {\delta}_{\mu \nu}
\label{K6}
\eeq
where the metric tensor $\theta_{IJ}$ is now diagonal and given
by
\beqa
 ({\theta}_{IJ})=
  \left( \begin{array}{ccc}
  {\delta}^{ab} & 0 & 0  \\
  0 &  {\delta}_{ij} & 0 \\
  0 & 0 & - \delta_{ij}    \end{array}  \right)
\label{K7}
\eeqa
Inserting the metric tensor (\ref{K7})
into (\ref{K6}) yields
\beqa
{\delta}_{ij} L_{\mu}\,^i L_{\nu}\,^j = {\delta}_{\mu \nu} +
{\delta}_{ij} {\tilde{L}}_{\mu}\,^i  {\tilde{L}}_{\nu}\,^j +
{\delta}^{ab} {\cal{L}}_{\mu a} {\cal{L}}_{\nu b}
\label{K8}
\eeqa
Next, let us introduce a set of complex coordinates
as follows \cite{Gil}.  The real matrix ${\Upsilon}_{\mu}\,^I$
has two columns and
r+2 rows. By combining the two real entries in each row into a single
complex variable
\beqa \phi_I = {\Upsilon}_1\,^I + i {\Upsilon}_2\,^I
\label{K9}
\eeqa
one arrives at a set of r+2 complex coordinates.  In terms of these
new complex variables the orthogonality relation (\ref{K6}) now reads
\beqa
   - {\phi}^{\dagger} I_{r,2} \phi &=& 2 \nonumber\\
   {\phi}^T I_{r,2} {\phi} &=& 0
\label{K10}
\eeqa
where $\phi$ is a complex column vector with complex entries ${\phi}_I$
and where $I_{r,2}$ denotes a diagonal matrix with entries
$I_{r,2} = diag(1,...,1,-1,-1)$.  This set of $r+2$ complex coordinates
is not yet a suitable one for constructing
K\"{a}hler potentials, since they are not unconstrained but rather satisfy
the orthogonality properties (\ref{K10}).  The next step then consists in
identifying a particular set of $r$ unconstrained complex coordinates
which the K\"{a}hler potential is going to
depend on.  It is known
\cite{FerKouLuZw}
how to find such a set of analytic coordinates for a general
$\SOd$ coset.  For concreteness and in order to keep the formulae
as simple as possible, we will in the following focus on the $\S4$
coset and explicitly construct an analytic set of coordinates
for it.  To proceed,
we first introduce an explicit parametrisation of the metric $G_{ij}$ of
the underlying $T_2$ torus as follows
\beqa
G_{ij} =
\left( \begin{array}{cc}
        R_1^2 & R_1R_2 \cos{\theta} \\
        R_1R_2 \cos{\theta}  & R_2^2
      \end{array}  \right)
\label{K11}
\eeqa
The associated zweibein $E_{\mu i}$ satisfying $G_{ij} = {\delta}^{\mu \nu}
E_{\mu i} E_{\nu j}$ and its inverse are then given by
\beqa
E_{\mu i} =
\left( \begin{array}{cc}
        R_1 & R_2 \cos{\theta} \\
        0 & R_2 \sin{\theta}
      \end{array}  \right), \,\,\,
E_{\mu}\,^ i =
\left( \begin{array}{cc}
        \frac{1}{R_1} & \frac{-1}{R_1} \frac{\cos{\theta}}{ \sin{\theta}} \\
        0& \frac{1}{R_2} \frac{1}{\sin{\theta}}
      \end{array}  \right)
\label{K12}
\eeqa
Then, by inserting parametrisation (\ref{K11}) and (\ref{K12}) into
(\ref{K9}),
the six complex
coordinates $\phi_I$ can be readily expressed in terms of the real moduli
fields $G_{ij}, B_{ij}$ and ${\cal{A}}_{ai}$ as
\beqa
\phi_1 &=& \frac{1}{\sqrt{Y}} \left( {\A}_{11} \frac{\sqrt{G}}{G_{11}}
+i(-{\A}_{11} \frac{G_{12}}{G_{11}} + {\cal{A}}_{12}) \right) \nonumber\\
\phi_2 &=& \frac{1}{\sqrt{Y}} \left( {\A}_{21} \frac{\sqrt{G}}{G_{11}}
+i(-{\A}_{21} \frac{G_{12}}{G_{11}}
+ {\cal{A}}_{22}) \right) \nonumber \\
\phi_3 &=& \frac{1}{\sqrt{Y}} \left( \sqrt{G} (1 - \frac{1}{G_{11}}
+ \frac{1}{4}
\frac{ {\cal{A}}^a\,_1 {\cal{A}}_{a1}}{G_{11}})
+i(-B_{12} + \frac{G_{12}}{G_{11}} (1- \frac{1}{4}
{\cal{A}}^a\,_1 {\cal{A}}_{a1}) + \frac{1}{4}
{\cal{A}}^a\,_1 {\cal{A}}_{a2}) \right) \nonumber\\
\phi_4 &=& \frac{1}{\sqrt{Y}} \left( \frac{\sqrt{G}}{G_{11}}(G_{12}
+ B_{12} + \frac{1}{4} {\A}^a\,_1 {\A}_{a2}) \right.\nonumber\\
&& \left. +i(-1+ \frac{G}{G_{11}} - B_{12} \frac{G_{12}}{G_{11}}
-\frac{1}{4} {\A}^a\,_1 {\A}_{a2} \frac{G_{12}}{G_{11}}+
\frac{1}{4} {\A}^a\,_2 {\A}_{a2}) \right) \nonumber \\
\phi_5 &=& \frac{1}{\sqrt{Y}} \left( \sqrt{G} (1 + \frac{1}{G_{11}}
+ \frac{1}{4} \frac{ {\cal{A}}^a\,_1 {\cal{A}}_{a1}}{G_{11}})
+i(-B_{12} - \frac{G_{12}}{G_{11}} (1+ \frac{1}{4}
{\cal{A}}^a\,_1 {\cal{A}}_{a1})+ \frac{1}{4}
{\cal{A}}^a\,_1 {\cal{A}}_{a2}) \right) \nonumber\\
\phi_6 &=&  \frac{1}{\sqrt{Y}} \left( \frac{\sqrt{G}}{G_{11}}(G_{12}
+ B_{12} + \frac{1}{4} {\A}^a\,_1 {\A}_{a2}) \right. \nonumber\\
&& \left. +i(1+ \frac{G}{G_{11}} - B_{12} \frac{G_{12}}{G_{11}}
-\frac{1}{4} {\A}^a\,_1 {\A}_{a2} \frac{G_{12}}{G_{11}} +
\frac{1}{4} {\A}^a\,_2 {\A}_{a2}) \right)
\label{K13}
\eeqa
where $G=\det{G_{ij}}$ and
\beq Y= 4 \frac{G}{G_{11}}
\label{K14}
\eeq
Note that there is
an overall factor $\sqrt{Y}$ appearing in front of each of the six $\phi_I$.
It is therefore convenient
\cite{FerKouLuZw} to introduce rescaled coordinates
\beqa
y_I = \sqrt{Y} \phi_I
\label{K15}
\eeqa
satisfying the rescaled constraints
\beqa
   - y^{\dagger} I_{4,2} \, y &=& 2 Y \nonumber\\
   y^T I_{4,2} \, y &=& 0
\label{K16}
\eeqa
The importance of the overall factor lays in that it determines the
K\"{a}hler potential $K$ as
\beqa K=-lnY
\label{K17}
\eeqa
provided one chooses a solution to the constraint equations
(\ref{K16}) possessing
an $SO(1,3)$ symmetry (more generally an $SO(1,r-1)$ symmetry in the case
of an $\SOd$ coset) \cite{FerKouLuZw}.
That is, by choosing four of the $y_I$ as unconstrained coordinates
one then seeks a solution of the constraint equations (\ref{K16})
for the two remaining
$y$-coordinates as well as for $Y$ which exhibits an $SO(1,3)$ symmetry.
We will, for concreteness,  choose
$y_1,y_2,y_3$ and $y_5$ as unconstrained variables.
Then, it can be checked that $Y$, given in (\ref{K14}), can be
expressed in terms of $y_1,y_2,y_3$ and $y_5$ as
\beqa
Y &=& \frac{1}{4} \left((y_5 + {\bar{y_5}})^2 - (y_1 + {\bar{y_1}})^2 -
(y_2 + {\bar{y_2}})^2 -
(y_3 + {\bar{y_3}})^2 )\right)
\label{K18}
\eeqa
Note that (\ref{K18})
exhibits an $SO(1,3)$ symmetry.  Inserting (\ref{K18}) into (\ref{K16})
yields
\beqa
 y_4^2 - y_6^2 &=& y_5^2 - y_1^2 - y_2^2 - y_3^2 \nonumber\\
 {\mid y_4 \mid}^2 - {\mid y_6 \mid}^2 &=& - \frac{1}{2} (
y_5^2 - y_1^2 - y_2^2 - y_3^2  +
{\bar y}_5^2 - {\bar y}_1^2 - {\bar y}_2^2 - {\bar y}_3^2 )
\label{K19}
\eeqa
 From (\ref{K13}), on the other hand, it follows that
\beq y_6 - y_4 = 2i
\label{K20}
\eeq
Then, it can be checked that
the following \cite{FerKouLuZw} solves (\ref{K19}) subject to (\ref{K20})
\beqa
y_4 &=& -i \left( 1 - \frac{1}{4}
( y_5^2 - y_1^2 - y_2^2 - y_3^2) \right) \nonumber\\
y_6 &=& i \left(1 + \frac{1}{4}(
y_5^2 - y_1^2 - y_2^2 - y_3^2) \right)
\label{K21}
\eeqa
Note that the solution (\ref{K21}) also
exhibits an $SO(1,3)$ symmetry.
Thus, (\ref{K18}) and (\ref{K21}) are a
solution to the orthogonality relations (\ref{K16}) with manifest $SO(1,3)$
symmetry.

The analytic structure
of the K\"{a}hler potential $K=-lnY$ can be made manifest \cite{FerKouLuZw}
by introducing
four complex fields $M_{ij}$ as
\beqa
(M_{ij})= \left( \begin{array}{cc}
        y_5 + y_3& y_1 - i y_2 \\
        y_1 + i y_2 & y_5 - y_3
      \end{array}  \right)
\label{K22}
\eeqa
Then, $Y$ is given by
\beqa
Y=\frac{1}{4} \det{(M_{ij}+{\bar{M}}_{ij})}
\label{K23}
\eeqa
Finally, introducing the linear combinations
\beqa
T&=&y_5+y_3 \nonumber\\
2U&=&y_5-y_3 \nonumber\\
B&=&y_1 - iy_2  \nonumber\\
C&=&y_1 + iy_2
\label{K24}
\eeqa
yields
\beqa
(M_{ij})= \left( \begin{array}{cc}
        T & B \\
        C & 2U
      \end{array}  \right)
\label{K25}
\eeqa
It follows from (\ref{K13}) that
the complex $T,U,B$ and $C$ moduli fields are expressed in terms of the
real ones as
\beqa
T&=& 2\left(\sqrt{G} (1+ \frac{1}{4}
\frac{ {\A}^a_1
{\A}_{a1}}{G_{11}})
-i(B_{12}+\frac{1}{4}
{\A}^a_1 {\A}_{a1} \frac{G_{12}}{G_{11}}
- \frac{1}{4} {\cal{A}}^a_1 {\cal{A}}_{a2} )
 \right) \nonumber\\
U&=& \frac{1}{G_{11}} (\sqrt{G} - i G_{12} )\nonumber\\
B&=& \frac{1}{G_{11}} \left(
{\cal{A}}_{11} \sqrt{G} - {\A}_{21} G_{12} + {\A}_{22} G_{11}+
i (-{\A}_{11} G_{12} + {\A}_{12} G_{11} - {\A}_{21} \sqrt{G})\right)
\nonumber\\
C&=&\frac{1}{G_{11}} \left(
{\cal{A}}_{11} \sqrt{G} +  {\A}_{21} G_{12} - {\A}_{22} G_{11} +
i (-{\A}_{11} G_{12} + {\A}_{12} G_{11} +{\A}_{21} \sqrt{G}) \right)
\label{K26}
\eeqa
The $T$ and the $U$ modulus are related to the geometrical data of the
two-dimensional torus $T_2$.  The $T$ modulus is associated with
deformations of the K\"{a}hler class.  It reduces to the well-known
expression \cite{Ver}
when turning off the real Wilson lines $\A_{ai}$.  The $U$ modulus
is associated with deformations of the complex structure.  Note that it
doesn't get admixtures of real Wilson lines $\A_{ai}$, that is, it remains
given by the
well-known expression \cite{Ver} when no Wilson lines are turned on.
Finally, the complex $B$ and $C$ moduli are linear expressions
in the real Wilson lines $\A_{ai}$. They vanish when turning off the real
Wilson lines $\A_{ai}$.  Thus, they qualify to be called complex Wilson lines.
The K\"{a}hler potential reads
\beqa
K &=& -lnY =- ln\left(\frac{1}{4}
\det{(M_{ij}+{\bar{M}}_{ij})}\right) \nonumber\\
&=&- ln \left( (T + {\bar{T}})(U + {\bar{U}})-
\frac{1}{2} (B + {\bar{C}})(C+\bar{B}) \right) + const
\label{K27}
\eeqa
A few remarks are at hand.  First note that
in the absence of Wilson lines ($B=C=0$) the K\"{a}hler potential
splits into the sum $K=K(T, {\bar{T}}) + K(U, \bar{U})$, which is the
well-known K\"{a}hler potential for the coset $\SO2 = \SU \otimes \SU$.
On the other
hand, turning on Wilson lines leads to the K\"{a}hler potential
(\ref{K27}) which
doesn't split into two pieces anymore.  This is so, because the $\S4$ coset
doesn't factorise anymore into two submanifolds.  Also note that the complex
Wilson lines $B$ and $C$ do not just give rise to
${\bar{B}}B$ and ${\bar{C}}C$ terms in the K\"{a}hler potential but also
to holomorphic $BC$ and antiholomorphic ${\bar{B}}{\bar{C}}$ pieces.
This will have important consequences when discussing target space duality
symmetries of the K\"{a}hler potential, as discussed in the next section.
Finally, let us point out that, even in the presence of Wilson lines, the
K\"{a}hler potential is in terms of the real moduli still given as
$K=-ln \frac{4G}{G_{11}}$ and thus still proportional to the volume
of the internal manifold.

We proceed with a discussion of the $\P3$ coset.  Inspection of (\ref{K5})
shows that a $\P3$ coset occurs when only retaining the first components
${\A}_{11}$ and ${\A}_{12}$ and setting all
the other components of the
two Wilson lines $\bf{A_1}$ and $\bf{A_2}$ to zero.  Then, it follows
from (\ref{K13}) that $\phi_2 = 0$ and, hence, $y_2 = 0$.  The corresponding
solutions (\ref{K18}) and (\ref{K21}) then exhibit an $SO(1,2)$ symmetry
and the K\"{a}hler potential is thus given again by $K=-ln Y$.  The complex
moduli can be read off from (\ref{K24}), where now $B=C=y_1$.  The
complex $T, U$ and $B$ moduli are expressed in terms of the real ones
as in (\ref{K26}) with ${\A}_{21} = {\A}_{22} = 0$.  It follows
from (\ref{K27}) that the K\"{a}hler potential for
the $\P3$ coset reads
\beqa
K &=& - ln \left( (T + {\bar{T}})(U + {\bar{U}})-
\frac{1}{2} (B + {\bar{B}})^2 \right) + const
\label{KK28}
\eeqa
This concludes the discussion of the $\S4$ and $\P3$ cosets.  Let us
point out again that any other $\SOd$ coset can be analysed along
very similar lines.

We now discuss the situation when the twist ${\theta}=({\theta}_i\,^j)$
acting on the internal 2-torus $T_2$ has eigenvalues
different from $-1$.  Introducing ${\theta}^{T,-1}=({\theta}^j\,_k)$
and
analysing the consistency conditions \cite{Jun,Moh2}
\beqa
G_{ij}\, {\theta}^j\,_k = {\theta}_i\,^j \, G_{jk} \nonumber\\
({\bf A}_i \cdot {\bf A}_j) \, {\theta}^j\,_k = {\theta}_i\,^j \,
({\bf A}_j \cdot {\bf A}_k)
\label{K28}
\eeqa
for the
internal metric $G_{ij}$ and for the matrix ${\bf A}_i \cdot {\bf A}_j$
shows that both $G_{ij}$ and ${\bf A}_i \cdot {\bf A}_j$
have only one independant
entry each.  Denoting these independant entries by $G_{11}$
and ${\bf A}_1 \cdot {\bf A}_1$, respectively, yields
\beqa
(G_{ij})= \frac{G_{11}}{2} \left( \begin{array}{cc}
        2 & \alpha  \\
        \alpha  & \beta
      \end{array}  \right),\;\;\;\;
({\bf A}_i \cdot {\bf A}_j)=
\frac{{\bf A}_1 \cdot {\bf A}_1}{2} \left( \begin{array}{cc}
        2 & \alpha  \\
        \alpha  & \beta
      \end{array}  \right)
\label{K29}
\eeqa
where $\alpha$ and $\beta$ are some twist dependent constants.
It then follows from (\ref{K26})
that the $U$-field takes a constant value given by
\beqa U = \frac{\sqrt{2 \beta - \alpha^2}}{2} - i \alpha
\label{K30}
\eeqa
The $T$-field survives as a modulus
and is given by
\beqa T=2 \left( \frac{\sqrt{2 \beta - \alpha^2}}{2}
( G_{11} + \frac{1}{4} {\A}^a\,_1 {\A}_{a1} ) - i B_{12}  \right)
\label{K31}
\eeqa
Note that only the first real Wilson line enters in this expression.
In fact, the second real Wilson line is not independent anymore, but
rather fixed in terms of the first real Wilson line \cite{IbaNilQue}.
This follows
directly from the consistency condition \cite{Moh2}
\beqa
\theta' {\bf A}_i = A^A\,_i \, {\theta'}_A\,^B \, {\bf e}_B = {\theta}_i\,^j
\, {\bf A}_j
\label{K32}
\eeqa
for the continuous Wilson lines.  Then, (\ref{K32}) yields
\beqa
{\bf A}_1 \rightarrow {\bf A}_2 = {\theta}_1\,^j \, {\bf A}_j =
{\theta'} \, {\bf A}_1 = A^A\,_1 \, {\theta'}_A\,^B \, {\bf e}_B
\label{K33}
\eeqa
which, indeed, expresses ${\bf A}_2$ in terms of ${\bf A}_1$.

It is useful to introduce a complex Wilson
line as
\beqa
\A = {\A}_{11} + i {\A}_{21}
\label{K34}
\eeqa
Then, it follows from (\ref{K26}) that
the combination $B+ \bar{C}$ can be written as
\beq
B + \bar{C} = \sqrt{ 2 \beta - \alpha^2} \bar{\A}
\label{K35}
\eeq
and the K\"{a}hler potential (\ref{K27}) as
\beqa
K = - ln \left( T + \bar{T} - \frac{\sqrt{2 \beta - \alpha^2}}{2}
\bar{\A} \A \right) + const
\label{K36}
\eeqa
This is the K\"{a}hler potential for the $\SK$ coset. More generally,
a $\SUS$ coset will be parametrised by one complex $T$ modulus and
n-1 complex Wilson lines $\A_l$ given by
\beqa T &=& 2 \left( \frac{\sqrt{2 \beta - \alpha^2}}{2}
( G_{11} + \frac{1}{4} {\A}^a\,_l {\A}_{al} ) - i  B_{12}\right) \nonumber \\
\A_l &=& A_{1l}+i{\A}_{2l}
\label{K37}
\eeqa
The corresponding K\"{a}hler potential reads
\beqa
K = - ln \left( T + \bar{T} - \frac{\sqrt{2 \beta - \alpha^2}}{2}
\bar{\A}_l {\A}_l \right) + const
\label{K38}
\eeqa
Finally, let us point out that additional untwisted matter fields,
charged or uncharged under the generic gauge group
which survives when turning on Wilson lines,
enter into the K\"{a}hler potentials (\ref{K27}) and (\ref{K38})
in precisely the same way as the complex Wilson lines $B$, $C$,
${\cal A}_l$.  Hence, the modular symmetry properties of these
K\"{a}hler potentials are preserved by the inclusion of untwisted matter
fields.

\setcounter{equation}{0}

\section{Examples of Target Space Duality Symmetries}

\hspace*{.3in}
We are now poised to discuss the symmetry properties of the
K\"{a}hler potentials we constructed in the previous section.
We will, in particular, be concerned with target space duality
symmetries, also referred to as modular symmetries.  As stated in section 4,
the spectrum
of untwisted states of an orbifold theory is invariant under
certain discrete transformations of the winding and momentum numbers
accompanied by discrete transformations of the moduli fields.
These transformations of the moduli fields induce particular
K\"{a}hler transformations of the K\"{a}hler potential and, thus, are
symmetries of the tree-level low-energy Lagrangian describing the
coupling of moduli fields to supergravity.

As explained in section 4,
modular transformations act on the
vector $v^T=(q^A,n^i,m_i)$ of quantum numbers
as integer valued transformations $\Omega$
\beqa v^{\prime} = \Omega^{-1} v
\label{E1}
\eeqa
As discussed earlier,
$\Omega$ must satisfy
(\ref{M8}).
Modular transformations (\ref{E1})
act on the real moduli matrix $\Sigma$ given in (\ref{M13})
as
\beqa \Sigma \rightarrow \Omega^T \Sigma \Omega
\label{E2}
\eeqa

We begin by discussing modular symmetries of the $\SOd$ cosets.
For concreteness, we will again focus on the $\S4$ coset.
The associated modular group $G_{O}$ will be called $O(4,2;{\bf Z})$.
It is crucial at this point to notice that some care is required
in order
to specify this group.
If the sixdimensional sublattice $\Gamma_{4;2}$ of the Narain lattice
$\Gamma_{22;6}$, on which $G_{O}$ acts,
happens to
factorize (that is, if
there is an orthogonal direct decomposition of $\Gamma_{22;6}$ into
$\Gamma_{4;2}$ and its complement) then $G_{O}$ will be the
group given by
\be   \{ M \in Gl(6; {\bf Z}) | M^{T} H M = H \}
\eq
where $H$ is the lattice metric of $\Gamma_{4;2}$.
But such a decomposition will in general not exist \cite{ErlSpa} and therefore
one has the further constraint that the elements
of $G_{O}$ must also act cristallographically on the full lattice
$\Gamma_{22;6}$. The resulting constraints should be similar to those
found in the case where the internal six--dimensional torus does not
factorize
\cite{BLST1}.

For definiteness and simplicity, we will only consider the following
case which is the simplest one.  As already explained in the
last section, we demand that the internal torus decomposes as
$T_{4} \oplus T_{2}$, where the twist $\theta$ acts on $T_{2}$ as
$-I$. The corresponding directions are labeled by $i=1,2$.
The modular group $G_{O}$ then contains the well known
group
\be O(2,2;{\bf Z}) = \{ M \in Gl(4, {\bf Z}) | M^{T}
     \eta M = \eta \}
\eq
as a subgroup, where
\[
     \eta = \left( \begin{array}{cc}
     0 & I \\ I & 0 \\
    \end{array} \right)
\]
A complete set of generators for $O(2,2,Z)$
can be found in \cite{GPR}.
$O(2,2,Z)$ acts non-trivially on the components $n^{i}$,
$m_{i}$ ($i = 1,2$) corresponding to the four basis vectors
$\ov{k}_{i}$, $k^{i}$ of the Narain lattice.
In order to find the full group $G_{O}$ we must identify those
quantum numbers $q^{A}$ which will transform non-trivially under
it. Since, by definition, $G_{O}$ is the group of modular
transformations in the $-1$ eigenspace of the twist $\Theta$, this then
means that we must identify all
basis vectors
$l_{A}$ of the Narain lattice which transform non-trivially under the
projection of the twist to the $-1$ eigenspace.
In \cite{Moh2} it was shown that,
in the absence of discrete background fields, the $l_A$ transform
under the twist
$\Theta$ with
the same integer valued matrix as the $E_{8} \otimes E_{8}$ basis
vectors ${\bf e}_{A}$. Note furthermore that, due to
the explicit
form (\ref{lA}) of the $l_{A}$,
modular transformations of the $q^{A}$ among
themselves and automorphisms of $E_{8} \otimes E_{8}$ are in a
one to one correspondence. Starting from these observations we
can find choices for the twist which are quite close to the situation
where the corresponding lattice factorizes. Namely, we will choose
the gauge twist $\theta'$ such that its two eigenvalues $-1$
come from a coxeter twist in an $A_{1} \otimes A_{1}$ sublattice.
(Note, however, that this is not the most general situation. A general
Coxeter twist of an subalgebra of $E_{8} \otimes E_{8}$,
which may be used to define an $E_{8} \otimes E_{8}$
automorphism, will have several different eigenvalues.)

In order
to construct the modular group $G_{O}$ we proceed in two steps.
First consider the sixdimensional sublattice $\Gamma_{4;2}$
of $\Gamma_{22;6}$, which is spanned by $l_{i}$, $\ov{k}_{i}$,
$k^{i}$ ($i=1,2$), where $l_{1}$ and $l_{2}$ correspond to
the $A_{1} \otimes A_{1}$ sublattice. The group of pseudo--orthogonal
automorphisms of $\Gamma_{4;2}$ is then given by
\be  \{ M \in Gl(6; {\bf Z}) | M^{T} H M = H \}
\label{MHM}
\eq
with
\be  H = \left( \begin{array}{ccc}
         C & 0 & 0 \\
         0 & 0 & I \\
         0 & I & 0 \\
         \end{array} \right)
\eq
where $C =2 $ diag$(1,1)$ is the Cartan matrix of
$A_{1} \otimes A_{1}$.

In a second step we should then identify all elements of this group
which also act cristallographically on the full lattice.
There are three classes of elements which automatically fulfill
this condition, namely all elements of the subgroup $O(2,2;{\bf Z})$,
all Weyl automorphisms of $A_{1} \otimes A_{1}$ and shifts of
the $q^{i}$ by multiples of $n^{j}$, $m_{j}$ with
$i,j =1,2$. To be able to say something about other elements
would, however, require a more detailed analysis. Therefore we will,
in the following,
only consider various particularly interesting elements belonging
to these three classes, and we will
work out the corresponding transformation properties
of the real and complex moduli
as well as
of the K\"ahler potential.

There are,
actually, three ways of deriving the action of the modular group $G_O$
on the real moduli
fields $G_{ij}, B_{ij}$ and $A_{Ai}$.  We will, in the following, make
use of all three of them.  They are as follows.
The transformation law of the
real moduli
can, in principle, be obtained
from (\ref{E2}).  This, however, can prove to be quite cumbersome.
An alternative way of obtaining the transformation law of the real moduli
fields is given by looking at the transformation law of the
projective
coset
coordinate $Z$ given in (\ref{Pro}).
Modular
transformations (\ref{E1}) act on $Z$ by fractional
linear transformations (\ref{Tra}).
Yet another
way for deriving the transformation law of the real moduli
fields is to look at
the background field matrix $\varepsilon$ \cite{Ven,GPR} given by
\beqa
\varepsilon =
 \left( \begin{array}{cc}
     E & F \\
     0 & X \\
      \end{array}  \right)
=
 \left( \begin{array}{cc}
 2 ( G+B+\frac{1}{4} {\A}^a {\A}_a )_{ij} & A_{Ai}\\
      0 &  (G + B)_{AB} \\
      \end{array}  \right)
\label{E3}
\eeqa
where the $(G+B)_{AB}$-data on the $E_8 \otimes E_8$ root lattice are given by
\beq
(G+B)_{AB}=C_{AB},\,\,\, A > B,\,\,\,
(G+B)_{AA}=\frac{1}{2}C_{AA},\,\,\,  (G+B)_{AB}=0,\, A < B \eeq
Consider an element $\hat{g} \in O(4,4,Z)$ and the bilinear form
${\hat{\eta}}$
\beqa \hat{g} =
 \left( \begin{array}{cc}
     \hat{a} &  \hat{b}\\
     \hat{c}  & \hat{d} \\
      \end{array}  \right) \;\;\;,\;\;\;
\hat{{\eta}} =
 \left( \begin{array}{cc}
     0 & I\\
     I & 0 \\
      \end{array}  \right)
\label{E4}
\eeqa
where $\hat{a}, \hat{b}, \hat{c}, \hat{d}, I$ are $4 \times 4$-dimensional
matrices.  $\hat{g}$ satisfies
$\hat{g}^T \hat{\eta} \hat{g} = \hat{\eta}$.
The action of $O(4,4,Z)$ on $\varepsilon$ is
given as \cite{GivRo,GPR}
\beqa
\varepsilon^{\prime} = \hat{g}( \varepsilon )=
( \hat{a} \varepsilon + \hat{b}) (\hat{c} \varepsilon + \hat{d})^{-1}
\label{E5}
\eeqa
Then, the modular group $O(4,2,Z)$ is the subgroup of $O(4,4,Z)$ that
preserves the heterotic structure of $\varepsilon$ in (\ref{E3}) while acting
on $\varepsilon$ by fractional linear transformations (\ref{E5}).

The modular group $O(4,2,Z)$ contains an $O(2,2,Z)$ subgroup.  There is
a natural embedding \cite{GivRo,GPR} of $O(2,2,Z)$ into $O(4,2,Z)$ given
as follows.  Consider an element $g \in O(2,2,Z)$ and the bilinear
form $\eta$
\beqa
g =
 \left( \begin{array}{cc}
     a &  b \\
     c  & d \\
      \end{array}  \right) \;\;\;,\;\;\;
\eta =
 \left( \begin{array}{cc}
     0 & I\\
     I & 0 \\
      \end{array}  \right)
\label{E6}
\eeqa
where $a,b,c,d,I$ are $2 \times 2$-dimensional matrices.
$g$ satisfies $g^T \eta g = \eta$.
Then, the embedding of $O(2,2,Z)$ into $O(4,2,Z)$
is given as
\beqa
 \hat{a} =
 \left( \begin{array}{cc}
     \hat{a} &  0\\
       0     &  I  \\
      \end{array}  \right),\, \,\,
 \hat{b} =
 \left( \begin{array}{cc}
     b &  0\\
     0 &  0 \\
      \end{array}  \right),\, \,\,
 \hat{c} =
 \left( \begin{array}{cc}
    c & 0\\
    0 & 0 \\
      \end{array}  \right),\, \,\,
 \hat{d} =
 \left( \begin{array}{cc}
     d &  0 \\
     0 &  I \\
      \end{array}  \right)
\label{E7}
\eeqa
The action of $O(2,2,Z)$ on $\varepsilon$ yields
\beqa {\varepsilon}^{\prime} = \hat{g}(\varepsilon)=
 \left( \begin{array}{cc}
    E^{\prime} & (a-E^{\prime}c)F \\
    0 & X \\
      \end{array}  \right)
\label{E8}
\eeqa
where
\beqa
E^{\prime}=(aE+b)(cE+d)^{-1}
\label{E9}
\eeqa

Let us now look at the subgroup of $O(2,2,Z)$ modular transformations.
A set of generators for $O(2,2,Z)$ can be found in the literature \cite{GPR}.
Here, we will, in  the following, look at
specific $O(2,2,Z)$ modular transformations
and derive the transformation laws for the real moduli fields.
Consider the inverse duality transformation given by
\beqa
\Omega =
\left( \begin{array}{cc}
     I &  0 \\
     0 &  g^T \\
      \end{array}  \right) ,\,\,\,\,\,
g=
\left( \begin{array}{cc}
     0 & I \\
     I & 0  \\
      \end{array}  \right)
\label{E10}
\eeqa
It acts on the quantum numbers as in (\ref{E1}), thus interchanging
the winding numbers $n^i$ with the momentum numbers $m_i$.
 From (\ref{E2})
one finds that
\cite{Ven}
\beqa
G_{ij} &\rightarrow& \frac{1}{4}
\left( (G+B+\frac{1}{4} {\A}^a {\A}_a)^{-1} G
(G-B+\frac{1}{4} {\A}^a {\A}_a)^{-1} \right)_{ij} \nonumber\\
B_{ij} &\rightarrow&    - \frac{1}{4}
\left( (G+B+\frac{1}{4} {\A}^a {\A}_a)^{-1} B
(G-B+\frac{1}{4} {\A}^a {\A}_a)^{-1} \right)_{ij} \nonumber\\
{\A}^a\,_i {\A}_{aj} &\rightarrow& \frac{1}{4}
\left( (G+B+\frac{1}{4} {\A}^a {\A}_a)^{-1} {\A}^b {\A}_b
(G-B+\frac{1}{4} {\A}^a {\A}_a)^{-1} \right)_{ij}
\label{E11}
\eeqa
 From (\ref{E11}) it then follows that
\beq E \rightarrow \frac{1}{4} E^{-1}
\label{E12}
\eeq
which is in agreement with what one obtains from (\ref{E9}).
The transformation law of the $A_{Ai}$ can, alternatively,
be obtained from (\ref{M31}) in a straightforward way.  It is
consistent with the transformation property of the ${\cal A}^a\,_i
{\cal A}_{aj}$ given in (\ref{E11}).

We proceed to show that inverse duality is a symmetry
transformation of the K\"{a}hler potential.
To do so, one has to compute the transformation laws of the
complex moduli fields $T,U,B$ and $C$ given in (\ref{K26}).
We will, in the following, only list a few of the lengthy
expressions arising when working out (\ref{E11}) and
(\ref{M31}) explicitly.
For instance, it can be checked
that
\beqa
G_{11} &\rightarrow& \frac{1}{4} \frac{
G_{22}}{(\det (G+B+\frac{1}{4} {\A}^a {\A}_a )_{ij})^2}
\left(
(B_{12} + \frac{1}{4} {\A}^a\,_1 {\A}_{a2} - {\A}^a\,_2 {\A}_{a2}
\frac{G_{12}}{G_{22}})^2 \right. \nonumber\\
&+&  G (1 + \frac{1}{4}
\left. \frac{{\A}^a\,_2 {\A}_{a2}}{G_{22}} )^2 \right)
\label{E13}
\eeqa
and that
\beqa
G &\rightarrow& \frac{1}{16} \frac{G}{(\det (G+B+\frac{1}{4}
{\A}^a {\A}_a )_{ij})^2}
\label{E14}
\eeqa
It can also be verified that (\ref{E13}) can be rewritten as
\beqa
G_{11} \rightarrow \frac{1}{16} G_{11}
\frac{1}{(\det (G+B+\frac{1}{4} {\A}^a {\A}_a)_{ij})^2}
\mid U \mid^2
\mid T - \frac{1}{2} \frac{BC}{U} \mid^2
\label{E15}
\eeqa
Similarly, one finds from (\ref{M31}) that
\beqa
A_{11} &\rightarrow& - \frac{1}{2}
\frac{1}{({\det}(G-B+ \frac{1}{4} {\cal A}^a {\cal A}_a)_{ij})}
\left(A_{11} (G_{22}+\frac{1}{4} {\cal A}^a\,_2 {\cal A}_{a2}) \right.
\nonumber\\
&& \left.
- A_{12} (G_{12}+B_{12}+\frac{1}{4} {\cal A}^a\,_1 {\cal A}_{a2}\right)
\label{E16}
\eeqa
Inserting all these expressions into (\ref{K26})
yields
\beqa
U &\rightarrow&- \frac{T}{- UT + \frac{1}{2} BC} \nonumber\\
T &\rightarrow&-  \frac{U}{- UT + \frac{1}{2} BC} \nonumber\\
B &\rightarrow& \frac{B}{- UT + \frac{1}{2} BC} \nonumber\\
C &\rightarrow& \frac{C}{- UT + \frac{1}{2} BC}
\label{E17}
\eeqa
Note that, in the presence of the complex Wilson lines $B$ and $C$,
the $T$ and $U$ moduli now mix under inverse duality.
When
switching off the complex Wilson lines $B$ and $C$, no
mixing occurs and
one obtains the familiar transformation law for the $T$ and $U$ moduli
\cite{GPR}.
Finally, inserting (\ref{E17}) into (\ref{K27}) yields
\beqa
K \rightarrow K+F+\bar{F}
\label{E18}
\eeqa
with the holomorphic $F$ given by
\beqa F = ln \left( UT (1-\frac{1}{2} \frac{BC}{UT}) \right)
\label{E19}
\eeqa
A useful check on (\ref{E19}) is
to look
at how $Y=\frac{4G}{G_{11}}$ transforms under (\ref{E11}).
It follows from (\ref{E14}) and (\ref{E15}) that
\beqa
Y \rightarrow Y \frac{1}{|U|^2}
\frac{1}{\mid T - \frac{1}{2} \frac{BC}{U}\mid^2}
\label{E20}
\eeqa
Thus, $K= - \ln \frac{4G}{G_{11}}$ transforms indeed as
in (\ref{E18}) and (\ref{E19}).

Next, let us look at the subgroup of $O(2,2,Z)$ transformations which acts
as $SL(2,Z)_U$ transformations on the $U$ modulus.  Let
\beqa
\Omega =
\left( \begin{array}{cc}
     I &  0 \\
     0 &  g^T \\
      \end{array}  \right),\,\, \,\,
g^T = \left( \begin{array}{cc}
     A&   \\
      & A^{T,-1} \\
      \end{array}  \right)
\label{E21}
\eeqa
where $A \in SL(2,Z)$.  $SL(2,Z)$ is generated by two elements \cite{GPR}
\beqa
T=
\left( \begin{array}{cc}
     1 &  1 \\
     0 &  1 \\
      \end{array}  \right),\,\, \,\,
S=
\left( \begin{array}{cc}
     0 &  1 \\
    -1 &  0\\
      \end{array}  \right) \eeqa
Consider the case where $A=T^p$ with $p \in Z$.  Then,
the transformation law of the real moduli fields is readily obtained from
(\ref{E9}) and reads
\beqa
(G_{ij}) &\rightarrow& (G_{ij}) +
\left( \begin{array}{cc}
     1  &  pG_{11} \\
    pG_{11} &  p^2 G_{11}+2 p G_{11} \\
      \end{array}  \right) \nonumber \\
B_{12} &\rightarrow& B_{12} \nonumber\\
{\A}_{a1} &\rightarrow& {\A}_{a1} \nonumber\\
{\A}_{a2} &\rightarrow& {\A}_{a2} + p{\A}_{a1}
\label{E23}
\eeqa
Inserting (\ref{E23}) into (\ref{K26}) yields
\beqa
U &\rightarrow& U -i p \nonumber\\
T &\rightarrow& T \nonumber\\
B &\rightarrow& B \nonumber\\
C &\rightarrow& C \eeqa
and, hence,
\beqa K \rightarrow K \eeqa
Now, consider the case where
$A=S$.
Then, it follows from (\ref{E9}) that
\beqa
G_{11} &\leftrightarrow& G_{22} \nonumber\\
G_{12} &\rightarrow& - G_{12}  \nonumber\\
B_{12} &\rightarrow& B_{12}  \nonumber \\
{\A}^a\,_1 &\rightarrow& - {\A}^a\,_2 \nonumber\\
{\A}^a\,_2 &\rightarrow& {\A}^a\,_1
\label{E26}
\eeqa
Inserting (\ref{E26}) into (\ref{K26}) yields
\beqa
U &\rightarrow& \frac{1}{U} \nonumber\\
T &\rightarrow& T - \frac{1}{2} \frac{BC}{U} \nonumber\\
B &\rightarrow& - \frac{B}{i U} \nonumber\\
C &\rightarrow& -  \frac{C}{i U}
\label{E27}
\eeqa
Note the peculiar Wilson line admixture in the transformation law of
the $T$ modulus.  Its presence is necessary to make the K\"{a}hler
potential transform properly.  Indeed, inserting (\ref{E27}) into
(\ref{K27}) yields
\beqa
K &\rightarrow& K + F(U) + \bar{F} (\bar{U}) \eeqa
where
\beq F(U) = ln \,U \eeq
Under more general $SL(2,Z)_U$ transformations \cite{BLST2}
\beqa
A =
\left( \begin{array}{cc}
     \delta & \beta \\
      \gamma & \alpha \\
      \end{array}  \right),\,\,\,\,
\alpha \delta - \beta \gamma = 1 \eeqa
it can be checked that the following holds
\beqa
U &\rightarrow& \frac{\alpha U -i \beta}{i \gamma U + \delta} \nonumber\\
T &\rightarrow& T - \frac{i \gamma}{2} \frac{BC}{i
\gamma U + \delta} \nonumber\\
B &\rightarrow&  \frac{B}{i \gamma U + \delta} \nonumber\\
C &\rightarrow& \frac{C}{i \gamma U + \delta}
\label{E31}
\eeqa
and that
\beq
F(U) = ln(i \gamma U + \delta)
\label{FU}
\eeq

As it is well known \cite{GPR}, there is also a subgroup of $O(2,2,Z)$
transformations which act as $SL(2,Z)_T$ transformations on the
$T$ modulus.  Let \cite{BLST2}
\beqa \Omega=
\left( \begin{array}{cc}
      I & \\
        & g^T \\
      \end{array}  \right),\,\,\,\,
g^T=
\left( \begin{array}{cc}
     \alpha I & \gamma J \\
     -  \beta J& \delta I \\
      \end{array}  \right),  \,\,\,\, \alpha \delta -\beta \gamma = 1
\label{E32}
\eeqa
Then, similarly, it can be shown that
\beqa
T &\rightarrow& \frac{\alpha T -i \beta}{i \gamma T + \delta} \nonumber\\
U &\rightarrow& U - \frac{i \gamma}{2} \frac{BC}{
i \gamma T + \delta} \nonumber\\
B &\rightarrow&  \frac{B}{i \gamma T + \delta} \nonumber\\
C &\rightarrow&  \frac{C}{i \gamma T + \delta}
\label{E34}
\eeqa
and
\beq K \rightarrow K + F(T) + \bar{F} (\bar{T})
\label{E35}
\eeq
where
\beq F(T) = ln(i\gamma T + \delta)
\label{E36}
\eeq

Next, let us look at elements of $O(4,2,Z)$ which are not in the
$O(2,2,Z)$ subgroup.  The biggest additional subgroup which commutes
with $O(2,2,Z)$ is given by the group of automorphisms of the
$A_1 \otimes A_1$ sublattice of the $E_8 \otimes E_8$ root lattice.
Since this subgroup acts trivially on the winding numbers $n^i$
and on the momentum numbers $m_i$, we will not be interested in it.
Consider, however, the following additional generator
of $O(4,2,Z)$, whose action on the quantum numbers is non-trivial, as
follows.
Consider the generator \cite{Lauer}
\beqa
W_L=
 \left( \begin{array}{cccccc}
     1 & 0 &  1 & 0 & 0 & 0 \\
     0 & 1 &  0 & 0 & 0 & 0 \\
     0 & 0 &  1 & 0 & 0 & 0 \\
     0 & 0 &  0 & 1 & 0 & 0  \\
     -2& 0 & -1 & 0 & 1 & 0 \\
     0 & 0 &  0 & 0 & 0 & 1  \\
      \end{array}  \right)
\eeqa
It satisfies (\ref{MHM}). Now, look at the action of
the group element
\beq \Omega = (W_L)^{p}\,\,\,\,, \,\, p \in Z \eeq
on the quantum numbers as given in (\ref{E1}).  It produces a shift
in the first component of $q$ (the momentum vector on the $E_8 \otimes E_8$
lattice)
\beqa
q^1 \rightarrow q^1 - p\, n^1
\label{E50}
\eeqa
by $p$ units of the winding number $n^1$.  The corresponding transformation
of the real moduli fields can be read off from the transformation properties
(\ref{M25}) and (\ref{Tra}) of the projective coordinate $Z$ given in
(\ref{Pro}).
One finds that
\beqa
G_{ij} &\rightarrow& G_{ij} \nonumber\\
B_{12} &\rightarrow& B_{12} - \frac{1}{4}\, p A_{12} \nonumber\\
A^1\,_1 &\rightarrow& A^1\,_1 + p \nonumber\\
A^2\,_1 &\rightarrow& A^2\,_1 \nonumber\\
A^1\,_2 &\rightarrow& A^1\,_2 \nonumber\\
A^2\,_2 &\rightarrow& A^2\,_2
\label{E51}
\eeqa
Thus, the shift given in (\ref{E50}) is
accompanied by a shift in the first component
of the first Wilson line.  Inserting (\ref{E51}) into (\ref{K26}) yields
\beqa
T &\rightarrow& T + \frac{p}{2} C_{11} \,U + \frac{p}{2} \sqrt{C_{11}}
(B + C) \nonumber\\
U &\rightarrow& U \nonumber\\
B &\rightarrow& B + p \sqrt{C_{11}} \,U \nonumber\\
C &\rightarrow& C + p \sqrt{C_{11}} \,U
\label{E61}
\eeqa
and, from (\ref{K27}),
\beq K \rightarrow K
\label{E100}
\eeq
We would like to point out that the associated group element
$\hat{g} \in O(4,2,Z)$ reproducing (\ref{E51}) via (\ref{E5})
can be constructed
and that it is given by (\ref{E4}) with
\beqa
\hat{a}&=&
\left( \begin{array}{cc}
\left( \begin{array}{cc}
             1 & 0 \\
             0 & 1 \\
                  \end{array}\right) & p \left( \begin{array}{cc}
                      1 & 0 \\
                      0 & 0 \\
                  \end{array}\right)  \\
\left( \begin{array}{cc}
             0 & 0 \\
             0 & 0 \\
                  \end{array}\right) &  \left( \begin{array}{cc}
             1 & 0 \\
             0 & 1 \\
                  \end{array}\right) \\
      \end{array}  \right) \;\;,\;\;
\hat{b}=
\left( \begin{array}{cc}
 -\frac{1}{2} p^2 C_{11} \left( \begin{array}{cc}
    1&0 \\
    0&0 \\
   \end{array}\right) & \frac{1}{2} p C_{11} \left( \begin{array}{cc}
                      1 & 0 \\
                      0 & 0 \\
                  \end{array}\right)  \\
 -\frac{1}{2} p C_{11} \left( \begin{array}{cc}
    1&0 \\
    0&0 \\
  \end{array}\right)  &
\left( \begin{array}{cc}
    0&0 \\
    0&0 \\
  \end{array}\right)  \\
      \end{array}  \right)  \nonumber\\
\hat{c}&=& 0 \;\;,\;\;
\hat{d} = \hat{a}^{T,-1}
\eeqa
Note that, since $C_{11}$ is even integer valued, $\hat{b}$ is also
integer valued.

We now briefly discuss the target-space duality symmetries of the
K\"{a}hler potential (\ref{KK28}) for a $\P3$ coset.  The
associated modular group $G_O$ is given by $O(3,2,Z)$.  Under
$O(3,2,Z)$ the complex moduli $T,U$ and $B$ transform as in
(\ref{E17}), (\ref{E31}), (\ref{E34}) and
(\ref{E61}) with $C=B$.  The associated K\"{a}hler potential (\ref{KK28})
then transforms as in (\ref{E19}), (\ref{FU}),
(\ref{E36}) and (\ref{E100}),
where again $C=B$.

Finally, let us turn to the target-space duality symmetries of the
K\"{a}hler potential (\ref{K38}) for an $\SUS$ coset.  It possesses an
$SL(2,Z)_T$ symmetry as in (\ref{E32}) with the $T$ modulus and
the complex Wilson lines ${\A}_l$ transforming as
\beqa
T &\rightarrow& \frac{\alpha T -i \beta}{i \gamma T + \delta} \nonumber\\
\A_l &\rightarrow& \frac{\A_l}{i \gamma T + \delta} \label{last}\eeqa
and the K\"{a}hler potential transforming
as in (\ref{E35}) and (\ref{E36}).
Let us finally remark that the modular transformation rules (\ref{E31}),
(\ref{E34}) and (\ref{last}) agree with previous results
\cite{FLT,IbaLu,IbaLu2} for untwisted matter fields.

\section{Conclusion}

\hspace{.3in} In this paper we showed that the local
structure of the untwisted moduli space
of asymmetric  $Z_N$ orbifolds is given by a product of $\SUm$ and
$\MGd$ cosets.  We then especialised to the case of $(0,2)$ symmetric
orbifold compactifications
with continuous Wilson lines.  For the case where the underlying
6-torus $T_6$ is given by a direct sum $T_4 \oplus T_2$ we showed that
interestingly enough, when the twist on the internal
torus lattice has eigenvalues
$-1$, there are holomorphic terms in the associated K\"{a}hler potential
describing the mixing of complex Wilson lines.  These terms deserve
further study since they were recently shown \cite{LouKap,Anto}
to be of the type which induce a mass term for Higgs particles of the
order of the gravitino mass once supergravity is spontaneously
broken.  We proceeded to identify the associated target space duality
symmetry groups and explicitly checked that they induce particular
K\"{a}hler transformations of the K\"{a}hler potentials.  In the
case where the twist on the internal torus lattice has eigenvalues
of $-1$, the associated $T$ and $U$ modulus were shown to
mix under target space duality transformations due to the presence
of the holomorphic mixing terms in the K\"{a}hler potential.
In more general terms, the discussed orbifold examples clearly show that
for (0,2) compactifications the moduli spaces of the moduli
corresponding to the deformations of the K\"{a}hler class
and of the complex structure respectively do not in general factorize like
in the (2,2) compactifications, and that they get mixed
by target space modular transformations.

Having
thus checked that these target space
symmetries are indeed symmetries of the 4D tree level low-energy Lagrangian,
it would be very interesting to know how these modular symmetries
manifest themselves in the string loop threshold corrections
\cite{Dix,AntNar,AntGav,May}.  There,
one expects to find that Wilson lines break some of the duality symmetries
and it would be interesting to find out to what subgroups they are broken
down.  Also, when turning on continuous Wilson lines, one generically
expects
to find smaller gauge groups than the ones present in $(2,2)$ symmetric
orbifold compactifications \cite{IbaLerLu}.  Let us point out that it
would be interesting to determine the generic
gauge groups occuring at generic points in the moduli space of the
$(0,2)$ models discussed in this paper.  Work along these lines is in
progress \cite{CarLuMoh}.  Finally, it would also be of interest to
extend the above investigations to the twisted sector.  On the one hand,
twisted moduli are important, because orbifolds can be smoothen out
into Calabi-Yau manifolds by assigning non-zero vevs to twisted moduli.
On the other hand, it has
recently been pointed out \cite{Sas} that twisted sectors
in asymmetric orbifolds may give rise to additional space-time
supercharges.


\begin{thebibliography}{99}
\bibitem{DHVW1}
        L. Dixon, J. Harvey, C. Vafa, and E. Witten,
        {\em Nucl. Phys.} {\bf B 261}
        (1985)
        678.
\bibitem{DHVW2}
        L. Dixon, J. Harvey, C. Vafa, and E. Witten,
        {\em Nucl. Phys. } {\bf B 274}
        (1986)
        285.
\bibitem{NSV1}
        K. Narain, M. Sarmadi and C. Vafa,
        {\em Nucl. Phys.} {\bf B 288} (1987) 551.
\bibitem{NSV2}
        K. Narain, M. Sarmadi and C. Vafa,
        {\em Nucl. Phys.} {\bf B 356} (1991) 163.
\bibitem{IbaNilQue2}
      L. E. Ib\'{a}\~{n}ez, H. P. Nilles and F. Quevedo,
      {\em Phys. Lett.} {\bf B 187} (1987) 25.
\bibitem{IbaNilQue}
         L. E. Ib\'{a}\~{n}ez, H. P. Nilles and F. Quevedo,
         {\em Phys Lett.} {\bf B 192} (1987) 332.
\bibitem{CreJul}
       E. Cremmer, B. Julia, J. Scherk, S. Ferrara, L. Girardello
        and P. van Nieuwenhuizen,
        {\em Nucl. Phys.} {\bf B 147} (1979) 105.
\bibitem{CreFer}
       E. Cremmer, S. Ferrara, L. Girardello and A. van Proyen,
       {\em Nucl. Phys.} {\bf B212} (1983) 413.
\bibitem{IbaMasNilQue}
         L. E. Ib\'{a}\~{n}ez, J. Mas, H. P.  Nilles
         and F. Quevedo,
         {\em Nucl.Phys. } {\bf B 301} (1988) 157.
 \bibitem{Moh2}
         T. Mohaupt,
        {\em MS--TPI} 93--09, {\em HEP--TH} 9310184.
 \bibitem{IbaLerLu}
        L. Ib\'{a}\~{n}ez, W. Lerche, D. L\"{u}st and
      S. Theisen,
       {\em Nucl. Phys.} {\bf B 352} (1991) 435.
  \bibitem{PorZwi}
      M. Porrati and F. Zwirner,
      {\em Nucl. Phys.} {\bf B 326} (1989) 162.
\bibitem{FerPor}
      S. Ferrara and M. Porrati,
      {\em Phys. Lett. } {\bf B 216} (1989) 216.
\bibitem{FerKouLuZw}
      S. Ferrara, C. Kounnas, D. L\"{u}st and
      F. Zwirner,
       {\em Nucl. Phys.} {\bf B 365} (1991) 431.
 \bibitem{LouKap}
       J. Louis and V. Kaplunovsky,
         {\em Phys. Lett.} {\bf B 306} (1993) 269.
\bibitem{Anto}
      I. Antoniadis, E. Gava, K. S. Narain and T. R. Taylor,
       manuscript in preparation.
\bibitem{FLST}
    S. Ferrara, D. L\"{u}st, A. Shapere and S. Theisen,
        {\em Phys. Lett.} {\bf B225} (1989) 363.
\bibitem{Kaplu} V. Kaplunovsky,
        {\em Nucl. Phys.} {\bf B307} (1988) 145.
\bibitem{Dix}
        L. Dixon, V. Kaplunovsky and J. Louis,
        {\em Nucl. Phys.} {\bf B 355} (1991) 649.
\bibitem{AntNar}
        I. Antoniadis, K. Narain and T. Taylor,
      {\em Phys. Lett.} {\bf B 267} (1991) 37.
\bibitem{AntGav}
       I. Antoniadis, E. Gava and K. Narain,
       {\em Nucl. Phys.} {\bf 383} (1992) 93.
\bibitem{DFKZ}
     J.P. Derendinger, S. Ferrara, C. Kounnas and F. Zwirner,
           {\em Nucl. Phys} {\bf 372} (1992) 145.
\bibitem{AELN}
    I. Antoniadis, J. Ellis, R. Lacaze and D.V. Nanopoulos,
         {\em Phys. Lett.} {\bf B268} (1991) 188.
\bibitem{ILR}
      L. Ib\'{a}nez, D. L\"{u}st and G. Ross,
          {\em Phys. Lett.} {\bf 272} (1991)  251.
 \bibitem{IbaLu}
       L. Ib\'{a}\~{n}ez and D. L\"{u}st,
       {\em Nucl. Phys.} {\bf B 382} (1992) 305.
 \bibitem{May}
          P. Mayr and S. Stieberger,
          {\em Nucl. Phys.} {\bf B407} (1993) 725.
  \bibitem{BLST1}
      D. Bailin, A. Love, W. A. Sabra and S. Thomas,
      {\em QMW--TH--}93/21, {\em SUSX--TH--}93/13,
      {\em HEP--TH} 9309133.
\bibitem{BLST2}
      D. Bailin, A. Love, W. A. Sabra and S. Thomas,
      {\em QMW--TH--}93/31, {\em SUSX--TH--}93/17,
      {\em HEP--TH} 9312122.
 \bibitem{Nar}
         K. S. Narain,
         {\em Phys. Lett.} {\bf 169 B} (1986) 41.
\bibitem{NSW}
         K. S. Narain, M. H. Sarmadi and E. Witten,
         {\em Nucl. Phys.} {\bf B 279} (1987) 369.
\bibitem{Gin}
        P. Ginsparg,
        {\em Phys. Rev. D} {\bf 35} (1987) 648.
 \bibitem{Moh1}
        T. Mohaupt,
        {\em Int. Jour. Mod. Phys.} {\bf A 8}(1993) 3529.
  \bibitem{ErlJunLau}
         J. Erler, D. Jungnickel and J. Lauer,
         {\em Phys. Rev. D} {\bf 45} (1992) 3651.
\bibitem{Jun}
         D. Jungnickel,
      {\em MPI-Ph}/92-27.
 \bibitem{FKP}
        S. Ferrara, C. Kounnas and M. Porrati,
        {\em Phys. Lett.} {\bf B181} (1986) 263.
 \bibitem{CveLouOvr}
      M. Cvetic, J. Louis and B. A. Ovrut,
      {\em Phys. Lett. B} {\bf 206} (1988) 227.
 \bibitem{Erl}
      J. Erler,
      {\em MPI--PH}/92-21.
  \bibitem{ErlKle}
J. Erler, A. Klemm,
       {\em Comm. Math. Phys.} {\bf 153} (1993) 579,
       {\em HEP--Th}--9207111.
\bibitem{KKKOT}
         Y. Katsuki, Y. Kawamura, T. Kobayashi, N. Ohtsubo and K.
         Tanioka,
         {\em Prog. Theor. Phys.} {\bf 82} (1989) 171.
 \bibitem{CarLuMoh}
        G. L. Cardoso, D. L\"{u}st and T. Mohaupt,
        in preparation.
\bibitem{HolMyh}
      T. J. Hollowood and R. G. Myhill, {\em Int. Jour. Mod. Phys.}
      {\bf A 3} (1988) 899.
 \bibitem{Gil}
      R. Gilmore,
      Lie Groups, Lie Algebras and some of Their Applications.
      Wiley--Interscience, New York, 1974.
 \bibitem{Spa}
      M. Spalinski,
      {\em Phys. Lett.} {\bf B 275} (1992) 47.
 \bibitem{Sas}
      T. Sasada,
      {\em KOBE--TH}--94--01, {\em HEP--TH} 9403037.
 \bibitem{FerFreSor}
      S. Ferrara, P. Fre and P. Soriani,
      {\em Class. Quant. Grav.} {\bf 9} (1992) 1649,
      {\em HEP--TH} 9204040.
 \bibitem{ErlSpa}
       J. Erler, M. Spalinski,
      {\em MPI-PH}/92-61, {\em TUM--TH--}147/92,
      {\em HEP--TH} 9208038.
 \bibitem{GPR}
      A. Giveon, M. Porrati and E. Rabinovici,
      {\em RI}--1--94, {\em NYU--TH}.94/01/01, {\em HEP--TH} 9401139.
 \bibitem{Ver}
       R. Dijkgraaf, H. Verlinde and E. Verlinde, proceedings
        of {\em Perspectives in String Theory}, Copenhagen, 1987.
  \bibitem{GivRo}
         A. Giveon and M. Rocek,
           {\em Nucl. Phys.} {\bf 380} (1992) 128.
  \bibitem{Ven}
        A. Giveon, E. Rabinovici and G. Veneziano,
          {\em Nucl. Phys.} {\bf B 322} (1989) 167.
  \bibitem{Lauer}
         J. Lauer, unpublished.
   \bibitem{FLT}
   S. Ferrara, D. L\"{u}st and S. Theisen,
   {\em Phys. Lett.} {\bf B242} (1990) 39.
   \bibitem{IbaLu2}
   L. Ib\'{a}nez and D. L\"{u}st,
   {\em Phys. Lett.} {\bf B302} (1993) 38.






\end{thebibliography}
\end{document}